\documentclass[epj]{svjour}
\usepackage{graphics}
\usepackage{amsmath}
\usepackage{amsfonts}
\begin{document}
\title{Glass models on Bethe lattices}
\author{O. Rivoire\inst{1} \and
G. Biroli\inst{2} \and 
O. C. Martin\inst{1} \and
M. M{\'e}zard\inst{1}
}                     
%
%
\institute{LPTMS, Universit{\'e} Paris-Sud, Orsay Cedex 91405 France \and
Service de Physique Th{\'e}orique CEA-Saclay,
Orme des Merisiers - 91191 Gif sur Yvette France}
\date{}
%
\abstract{
We consider ``lattice glass models'' in which each site can be
occupied by at most one particle, and any particle
may have at most $\ell $ occupied nearest neighbors. Using
the cavity method for locally tree-like lattices, we derive the phase diagram,
with a particular focus on the vitreous phase and the highest packing
limit. We also study the energy landscape via 
the configurational entropy, and discuss different
equilibrium glassy phases. Finally, we show that a kinetic freezing, 
depending on the particular dynamical rules chosen for the model, can prevent the
equilibrium glass transitions.
} 
\maketitle
\section{Introduction}
\label{intro}
The thermodynamics of vitreous materials is a long standing yet very
alive subject of study \cite{Struik78}. In spite of much research, it
remains unclear whether this ``amorphous'' state of matter can exist
in equilibrium; even if it cannot, probably the underlying crystalline
equilibrium phase is irrelevant for understanding the physics of
glassy systems. As experience has shown in other contexts, 
a lot of understanding can be gained by looking at models simple enough
to be analyzed but that retain the essential physics. 
Following this strategy, several recent works have focused on lattice
models for structural glasses.

Lattice gas models with hard core exclusion, i.e., with each site being
occupied by at most one particle, designed to reproduce the glass
phenomenology, fall into two distinct classes. A first class consists
of kinetically constrained models \cite{RitortSollich02} which have a
glassy behavior forced by a {\it dynamical constraint} on allowed
moves, but otherwise trivial equilibrium properties. An example is the
Kob-Andersen model \cite{KobAndersen93} where a
particle is allowed to move only if before and after its move it has
no more than some number $m$ neighboring particles. In this case the
slow dynamic displays several remarkable properties like for example
an avoided transition toward a cooperative ``super-Arrhenius'' dynamics \cite{ToninelliBiroli03}.
Physically, the kinetically constrained models are based on the assumption
that the glassy behavior is due to an increasing dynamical correlation
length whereas static correlations play no role. Thus, the possibility
of a thermodynamic glass transition is excluded from the beginning.

In contrast, the models we will discuss here belong to a second class
where the glassy features are generated through a {\it geometric
constraint} on allowed configurations. In this case a thermodynamic
equilibrium glass transition independent of the chosen local dynamical
rules may exist. Indeed, as we shall show in the following, it takes place for models on a Bethe lattice. 
Such models were first introduced in Ref.~\cite{BiroliMezard01} and some variants have been elaborated since then \cite{PicaTarzia02,WeigtHartmann03}.
 
In this paper, we study the lattice glass models on ``Bethe lattices'' which are
random graphs with a fixed connectivity. This kind of approach provides an approximation scheme for the lattice glasses on Euclidean lattices
having the same value of the connectivity, in the same way as the Bethe
approximation allows one to compute approximate phase diagrams of non frustrated
systems. Furthermore this approximation can be improved systematically, at
least in principle, by implementing  higher order cluster variational methods.
But this random graph study is also interesting for its own sake. In particular
we analyze in detail  the limit of diverging chemical potential. 
In this limit, one recovers an
optimization problem which is the lattice version of close-packing of spheres, a
problem that has challenged mathematicians for many decades \cite{Hales98} and
is still matter of debate today \cite{Szipiro03}.

With respect to their Euclidean counterparts, the lattice glasses on Bethe
lattices have one important difference: the existence of the random lattice,
even with fixed connectivity, forbids crystalline ordering. This loss is also
an advantage in that the absence of crystal phases makes it easier to study
the glass phase. Indeed, when the density of the system is increased, we find
a thermodynamical phase transition from a liquid phase to a glass phase. We
can determine the phase diagram analytically, focusing in particular on the
liquid to glass transition. In many systems this transition is 
of the ``random first order''
\cite{KirkpatrickThirumalai87a,KirkpatrickThirumalai87b,KirkpatrickThirumalai89,KirkpatrickWolynes87} type, also called ``one-step replica symmetry breaking''
\cite{MezardParisi87b}. On the Bethe lattice there are actually two transitions: when
increasing the density or chemical potential, one first finds a dynamical
transition in which ergodicity is broken, then a static phase transition. The
intermediate phase is such that the thermodynamic properties are those of the
liquid phase, in spite of the ergodicity breakdown. At the static transition
the entropy and energy are continuous, with a jump in specific heat. Therefore
the Bethe-lattice glass models provide clear solvable examples of a system of
interacting particles exhibiting the scenario for the glass transition which
was proposed in \cite{KirkpatrickThirumalai87a,KirkpatrickThirumalai87b,KirkpatrickThirumalai89,KirkpatrickWolynes87} by analogy with some spin glass systems. This
scenario is known to be a mean-field one, which does not take into account the
nucleation processes that can occur in Euclidean space. It is generally
believed that nucleation processes transform the dynamical transition into
some cross-over of the dynamics from a fast one to a slow, activated
relaxation \cite{KirkpatrickThirumalai87a,KirkpatrickThirumalai87b,KirkpatrickThirumalai89,KirkpatrickWolynes87,BouchaudCugliandolo98}. 
Whether the static transition survives in realistic system is
unknown so far. In this paper we will not discuss the
relevance and modifications of the mean field scenario when applied to finite
dimensional problems since we have nothing to add to existing speculations. 
Let us just mention that the lattice glasses provide the
best examples on which these issues can be addressed. The first step of such a
study is to have a detailed understanding of the mean field theory, and this is
what we do in the present paper.

Note that some of the results have appeared in Ref.~\cite{BiroliMezard01}; here we give a new and extensive discussion on the nature of the different possible equilibrium glassy phases.

On top of the thermodynamic study, we have also studied the dynamical arrest
of the lattice glasses on Bethe lattices. The dynamical arrest depends on the specific dynamical rule that is implemented. We show that this dynamical arrest is in
general unrelated to the energy landscape transitions found in the
thermodynamic approach. In particular, in some models, a dynamical arrest
takes place at a density {\it smaller} than the one of the dynamical glass
transition. In kinetically constrained problems, such mean field
arrest transitions are known to become crossovers in finite dimensional
systems.
The corresponding arrest behavior on our lattice glass models is not known yet.

The outline of this paper is as follows. In Sect.~\ref{sec:2}, we
introduce the models on lattices of arbitrary types and define the
relevant thermodynamics quantities. In Sect.~\ref{sec:3}, we address
the case of loop-less regular lattices, called Cayley trees when
surface sites are included and Bethe lattices when surface effects can be neglected. We show that low densities
correspond to a liquid phase described by the Bethe-Peierls
approximation, but that inhomogeneous phases must be present at higher
densities. However, the strong
boundary dependence of models on Cayley trees does not allow one to
define such dense phase in the interior. To overcome this problem,
we consider instead in Sect.~\ref{sec:4} random regular graphs; they
share with Cayley trees a local tree-like structure but are free of
surface effects and thus provide a natural generalization of Bethe
lattices adequate for dense phases. Our study on random graphs is
performed by means of the cavity method \cite{MezardParisi02} which
predicts for high densities a glassy phase. In Sect.~\ref{sec:5}, we
focus on the close-packing limit and discuss in detail the nature of
this glassy phase. Finally in Sect.~\ref{sec:6}, we comment on the differences
between the equilibrium glass transition of our models and the kinetic
transitions or dynamical arrests related to specific local dynamical rules. In particular
we show that for some models the dynamical arrest takes place {\it before} the equilibrium glass transition.

\section{Lattice Models}
\label{sec:2}

\subsection{Constraints on local arrangements}
\label{subsec:2:1}

When packing spheres in three dimensions, the preferred local ordering is 
icosahedral; this does not lead to a periodic crystalline
structure and is the source of frustration. To model this type of frustration, we forbid certain local arrangements of the
particles on the lattice. This can be done using
two or $n$-body interactions. We follow~\cite{BiroliMezard01}
and set the interaction energy to be infinite if
a particle has strictly more than $\ell$ particles
as nearest neighbors. The interactions thus act as
``geometric'' constraints.

Note that these geometric constraints are very similar
to the kinetic constraints of the Kob-Andersen model \cite{KobAndersen93}.
However, as we shall show, the fact that they are encoded in an energy function 
makes a big difference physically, in particular
for the thermodynamic behavior.

We work in the grand canonical ensemble and introduce
a chemical potential. All energies being zero or infinity,
temperature plays no role. The thermodynamics
for a given system (i.e., a given lattice)
is then defined by the grand canonical partition function
\begin{equation}\label{eq:Xi}
\Xi(\mu) = \sum_{n_1,\dots,n_N\in \{0,1\}}C_\ell(n_1,\dots,n_N) ~ e^{\mu\sum_{i=1}^Nn_i}.
\end{equation}
The dynamical variables are the 
site occupation values: $n_i=0$ if site $i$ is empty and $n_i=1$ if
a particle is on that site. We take all the particles to
be identical. In Eq.~(\ref{eq:Xi}),
$\mu$ is the chemical potential,  
and $C_\ell(n_1,\dots,n_N)$
implements all the geometrical constraints: it is the product 
of local constraints, one for each of the $N$ sites. The term for site $i$ is
\begin{equation}
\theta \left(\ell-n_i\sum_{j\in\mathcal{N}(i)}n_j\right)
\end{equation}
where $\mathcal{N}(i)$ denotes the set of neighbors of $i$, and $\theta(x)=0$ if $x<0$, $\theta(x)=1$ if $x\geq 0$.

For sake of concreteness and simplicity, we will focus in the core of the
paper on the $\ell=1$ case when each particle has at most one neighbor, deferring the general $\ell$ case to Appendix~A. Most numerical results will be given for the ``basic model'', noted BM, for which $\ell=1$ and $k=2$.

\subsection{Observables}
\label{subsec:2:2}

For a given ``lattice'' of $N$ sites and type (Euclidean, tree, random graph, \ldots), and for a given form of constraints (i.e., a value of $\ell$),
there is only one parameter, the chemical potential $\mu$.
It is useful to introduce the grand potential $\mu
\Omega(\mu)$ and its density $\omega(\mu)\equiv \Omega(\mu)/N$, so that
\begin{equation}
\Xi(\mu) = \exp \left[ - \mu \Omega (\mu) \right]=\exp \left[ - N\mu \omega (\mu) \right].
\end{equation}
The pressure is given by $p=-\mu\omega(\mu)$ and the particle density is
\begin{equation}
\rho(\mu)  \equiv \left \langle \frac{ \sum_i n_i }{N} \right \rangle = 
\frac{1}{N\Xi(\mu)} \frac{d \Xi(\mu)}{d \mu} = -\frac{d (\mu\omega)}{d \mu}=\frac{d p(\mu)}{d \mu}.
\end{equation}
Clearly, $\rho$ is an increasing function of $\mu$.
When $\mu \to -\infty$, the system becomes empty, 
a typical equilibrium configuration having almost
all $n_i=0$, so $\rho \to 0$. In the opposite limit,
$\mu \to +\infty$, there is in effect a strong
{\em penalty} for each vacancy, but the geometrical constraints prevent
all sites from being occupied; then $\rho$ has a maximum value
strictly less than $1$. When $N \to \infty$, this value
is expected to converge to a limit $\rho_{\infty}$.
Obviously, $\rho_{\infty}$ depends on the type of lattice
and on the parameter $\ell$.

We can define similarly the entropy density $s(\mu)$:
\begin{equation}
\label{eq:defentropy}
s(\mu)  \equiv \frac{\ln \Xi(\mu)}{N} - \mu \rho(\mu)=-\mu\omega(\mu)-\mu\rho(\mu)
\end{equation}
where again $s(\mu)$ should have a well defined thermodynamic limit.

Other physical quantities that we shall study include
susceptibilities associated with two-site connected correlation functions of the type
$\langle n_i n_j \rangle_c$. At low densities the $n_i$ have only short range correlations. When $\mu$ increases, $\rho$ also
increases; then the constraints become more important
and correlations grow. When the density
is close to $\rho_{\infty}$, the system will be very ``rigid'',
allowing few fluctuations in the local density. 
It is plausible that the susceptibility will diverge at some
critical value of $\mu$, separating a liquid phase at low $\mu$ from 
a denser phase at large $\mu$.

The nature of this dense phase, and of the transition, will depend on the lattice, on the boundary conditions,
and need not be associated
with a crystalline order. When it is not crystalline,
we want a statistical description of the dominant equilibrium
configurations.
In particular, if these configurations form clusters, 
it is of interest to estimate the number of such clusters. We will do this by computing the ``configurational entropy'' $\Sigma(\rho)$ (also called {\it complexity}) associated with the number of clusters (also called {\it states}) of configurations with a given density $\rho$.

\section{Models on Cayley trees}
\label{sec:3}

\subsection{Iteration equations}
\label{subsec:3:1}
We first consider glass models defined on regular {\it trees}, i.e. connected graphs with no loops and fixed connectivity. We distinguish {\it rooted trees} where a site called the {\it root} is connected to only $k$ neighbors while all the other sites (except for those at the surface, that is the leaves of the tree) have $k+1$ neighbors. A {\it Cayley tree} is obtained by connecting a site to the roots of $k+1$ rooted trees. 

\begin{figure}
\resizebox{0.5\textwidth}{!}{
\centering
\includegraphics{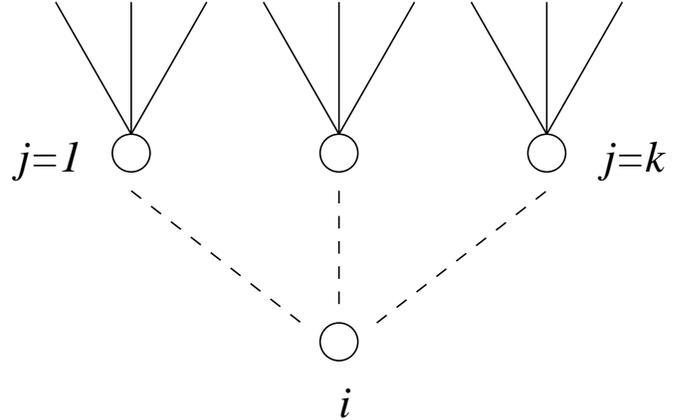}
}
\caption{An iterative method is used to compute the partition function on rooted trees. We begin with $k$ rooted trees with roots $j=1,\dots,k$ and form a new rooted tree by joining a site $i$ to each of them. The possible occupations of site $i$ depend on the occupations of the sites $j=1,\dots,k$ and on the type of constraint used.}
\label{fig:join}
\end{figure}

When the graph is a rooted tree, the grand canonical partition function
can be computed by recursion starting from the leaves. To do this, 
we follow conditional partition
functions because we need to know how to apply the constraints when joining the 
sub-trees (cf. Fig.~\ref{fig:join}). This provides a generalization of the well-known transfer matrix method for one dimensional systems (which can be viewed as rooted trees with $k=1$). 

In our class of constraints, we need to know
whether the root sites are occupied, and
if they are, whether they have $\ell$ occupied neighbors or less than that.
Let $\Xi_i^{(e)}$, $\Xi_i^{(u)}$ and 
$\Xi_i^{(s)}$ be the conditional partition functions for a rooted sub-tree
$i$ when its root node $i$ is
empty ({\it e}), occupied but the constraint unsaturated ({\it u})
and finally occupied and the constraint saturated ({\it s}),
i.e., the root site has $\ell$ neighboring particles. Then the
conditional partition functions for the rooted tree obtained by joining the
sub-trees
are easily computed. 

For instance, in the $\ell =1$ case where each particle can have at most one neighbor, when merging $k$ rooted trees
($j=1, \ldots k$) to obtain
a new tree rooted say at site $i$, we have
\begin{align}
\label{eq:xie}
\Xi^{(e)}_i&=\prod_{j=1}^k\left(\Xi^{(e)}_j+\Xi^{(u)}_j+\Xi^{(s)}_j\right),\\
\Xi^{(u)}_i&=e^\mu\prod_{j=1}^k\Xi^{(e)}_j,\\
\Xi^{(s)}_i&=e^\mu\sum_{j=1}^k\Xi^{(u)}_j\prod_{p\neq j}\Xi^{(e)}_p.
\label{eq:xis}
\end{align}
Naturally, the total partition function is the sum of
the conditional partition functions.

To study these recursions, it is convenient to consider
local fields defined via ratios of conditional partition functions.
Here we introduce on a root site $i$ two fields $a_i$ and $b_i$
defined as
\begin{align}
e^{-\mu a_i} &\equiv \frac{\Xi_i^{(e)}}{\Xi_i^{(e)} + \Xi_i^{(u)} + \Xi_i^{(s)} },\\
e^{-\mu b_i} &\equiv \frac{\Xi_i^{(e)}}{\Xi_i^{(e)} + \Xi_i^{(u)}}.
\end{align}
The first quantity is the probability for the root of a rooted tree to have
an empty site ({\it e}); the second is the ratio of that probability
and the probability to have a non saturated site.
The use of $\mu$ when defining these fields is to simplify the analysis at large
$\mu$ (cf. Sect.~\ref{sec:5}).
These two fields form a closed recursion under
the joining of sub-trees; for instance when $\ell=1$,
\label{eq:recursion1}
\begin{align}
\begin{split}
a_i&=\hat a (a_1,b_1,\dots,a_k,b_k)\\
&=\frac{1}{\mu}\ln \left[1+e^{\mu (1-\sum_{j=1}^ka_j)}\left(1+\sum_{j=1}^k(e^{\mu b_j}-1)\right)\right],
\end{split}\\
\label{eq:recursion2}
\begin{split}
b_i&=\hat b(a_1,b_1,\dots,a_k,b_k)\\
&=\frac{1}{\mu}\ln \left[1+e^{\mu (1-\sum_{j=1}^ka_j)}\right].
\end{split}
\end{align}
Note that the use of ratios of conditional partition
functions leads to  
recursions for two quantities rather than the initial three.
To recover all the information in the initial recursions,
we also keep track of the change in the grand potential. If
$\Omega_1$, \ldots $\Omega_k$ give the grand potentials of the $k$ sub-trees,
we have after the merging
\begin{equation}
\Omega_i = \sum_{j=1}^k \Omega_j + \Delta\hat \Omega_{\rm iter}(a_1,b_1,\dots,a_k,b_k)
\end{equation}
where $\Delta\hat \Omega_{\rm iter}(a_1,b_1,\dots,a_k,b_k)=\Delta\Omega_{\rm iter}$ is defined via
\begin{equation}
e^{-\mu\Delta\Omega_{\rm iter}}=\frac{\Xi_i}{\prod_{j=1}^k\Xi_j}=\frac{\Xi_i^{(e)}+\Xi_i^{(u)}+\Xi_i^{(s)}}{\prod_{j=1}^k\left(\Xi_j^{(e)}+\Xi_j^{(u)}+\Xi_j^{(s)}\right)}.
\end{equation}
With our definition of the fields, we have then $\Delta \hat\Omega_{\rm iter}=-\hat a$ .
From here on,  we shall use the
short-hand notation: $h\equiv (a,b)$ and $h_i=\hat h(h_1,\dots,h_k)$ with
\begin{equation}
\begin{split}
&\hat h\left(h_1=(a_1,b_1),\dots,h_k=(a_k,b_k)\right)=\\
&\left(\hat a(a_1,b_1,\dots,a_k,b_k),\hat b(a_1,b_1,\dots,a_k,b_k)\right).
\end{split}
\end{equation}

\subsection{Liquid phase}
\label{subsec:3:3}
Begin now on the leaves of a rooted tree, assuming that some
kind of boundary condition is specified there. For example, the 
$n_i$ could be fixed, or their probability distribution
could be given.
These determine the initial values of the conditional
partition functions and thus of the fields. We iterate the recursions,
propagating the fields away from the leaves by performing mergings.
When $\mu \ll -1$, these iterations are contracting
and so the fields converge to a value that is independent
of the starting values on the leaves. The distribution of fields in the bulk (away from the leaves) is then given by
\begin{equation}
\label{eq:distrs}
\mathcal{P}(h) = \delta(h-h_{\rm liq})
\end{equation}
with the fixed point condition $h_{\rm liq} = \hat h(h_{\rm liq}, \ldots ,h_{\rm liq})$.
We determine the fixed point for all $\mu$, and refer
to it as the {\it liquid} solution; its physical relevance includes at least the region $\mu \ll -1$.

We can now merge consistently $k+1$ rooted trees to obtain a Cayley
tree.  We then have a liquid (or paramagnetic) phase, all correlations
being short range and the heart of the Cayley tree being insensitive to the boundary
conditions, even though a finite fraction of the sites live on the
surface.  
In this regime, the homogeneous interior of the Cayley tree can be
used to define the {\it Bethe lattice model}.

In this context, also known as the {\it Bethe-Peierls approximation}, the density $\omega$ can be obtained from the following construction. Start with $(k+1)$ Bethe lattices; for each of them, pick an edge and remove it, leading to $2(k+1)$ infinite rooted trees. Then form {\it two} Bethe lattices by adding two sites and connecting each to $(k+1)$ of the rooted trees. Now the difference in grand potential between the resulting two Bethe lattices and the initial ones is just twice the grand potential per site (since two sites were added) and can be written as
\begin{equation}
\label{eq:betheomega}
2\omega=-(k+1)\Delta \Omega_{\rm edge}+2\Delta \Omega_{\rm site}
\end{equation}
where $\Delta \Omega_{\rm edge}$ is the difference in grand potential corresponding to adding an edge and $\Delta \Omega_{\rm site}$ to merging $(k+1)$ branches into a new site. 
Such quantities are easily expressed with the partition functions of
rooted trees; for instance for $\ell=1$ we obtain
\begin{align}
\begin{split}
\label{eq:Omega1}
&e^{-\mu\Delta \hat\Omega_{\rm site}(a_1,b_1,\dots,a_{k+1},b_{k+1})}\\
&=1+e^{\mu (1-\sum_{j=1}^{k+1}a_j)}\left(1+\sum_{j=1}^{k+1}(e^{\mu b_j}-1)\right)
\end{split}
\end{align}
and similarly
\begin{align}
\begin{split}
\label{eq:Omega2}
& e^{-\mu \Delta \hat\Omega_{\rm edge}(a_1,b_1,a_2,b_2)}\\
=&\left[\Xi^{(e)}_1\Xi^{(e)}_2+\Xi^{(e)}_1\left(\Xi^{(u)}_2+\Xi^{(s)}_2\right)+\left(\Xi^{(u)}_1+\Xi^{(s)}_1\right)\Xi^{(e)}_2\right.\\
&\left.+\Xi^{(u)}_1\Xi^{(u)}_2\right]\Big /(\Xi_1\Xi_2)\\
=&e^{-\mu(a_1+a_2)}\left( e^{\mu a_1}+e^{\mu a_2}+e^{\mu (b_1+b_2)}-e^{\mu b_1}-e^{\mu b_2}\right).
\end{split}
\end{align}
Note that we have the simple relation
\begin{equation}
\begin{split}
\Delta \hat\Omega_{\rm site}(h_1,\dots,h_{k+1})=&\Delta \hat\Omega_{\rm iter}(h_1,\dots,h_k)\\
&+\Delta \hat\Omega_{\rm edge}\left(\hat h(h_1,\dots,h_k),h_{k+1}\right)
\end{split}
\end{equation}
 whose interpretation is clear: the addition of one site by merging $k+1$ rooted trees $\mathcal{T}_1,\dots,\mathcal{T}_{k+1}$ can be decomposed into the construction of a new rooted tree $\mathcal{T}_0$ obtained by merging a site to the $k$ rooted trees $\mathcal{T}_1,\dots,\mathcal{T}_k$ and the addition of an edge between $\mathcal{T}_0$ and $\mathcal{T}_{k+1}$.
Moreover $\Delta \hat\Omega_{\rm site}$ is obtained from $\Delta
\hat\Omega_{\rm iter}$ by making the substitution $k\to k+1$. The expressions (\ref{eq:Omega1}-\ref{eq:Omega2})
are written for the general inhomogeneous case but simplify in the liquid phase where all the fields take their liquid value. 

\begin{figure}
\resizebox{0.5\textwidth}{!}{
\centering
\rotatebox{-90}{
\includegraphics{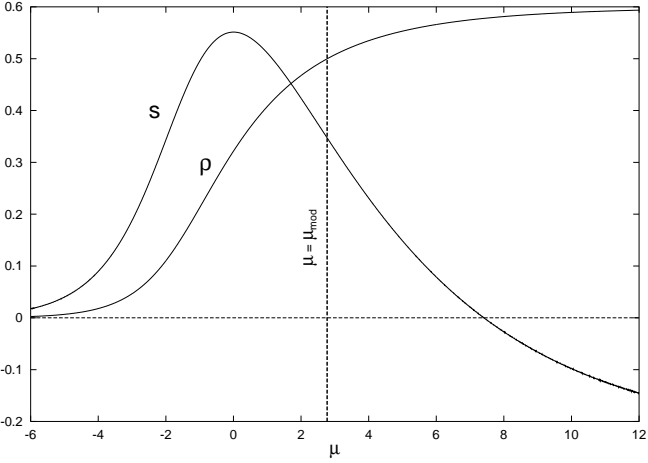}
}
}
\caption{$\mu$ dependence of the particle density $\rho$ and entropy $s$ of the liquid phase for the BM ($\ell=1$, $k=2$). The failure of the liquid phase to correctly describe the model at high $\mu$ is evidenced by the negative sign of the entropy for $\mu>\mu_{s=0}\simeq 7.40$. But in fact the liquid solution becomes linearly unstable well before, at $\mu_{\rm mod}\simeq 2.77$ (vertical line).}
\label{fig:liquid}
\end{figure}

In Fig.~\ref{fig:liquid}, we show as illustration the liquid's density $\rho$ and its entropy density $s$, as a function of $\mu$ for the BM.
Since the models are discrete, the equilibrium entropy should never
go negative. Nevertheless, the liquid solution at large $\mu$, $\mu>\mu_{s=0}$,
leads to $s_{\rm liq} < 0$ except in very special cases such as
$k=1$. Thus there must exist a phase transition at $\mu_c\leq \mu_{s=0}$, and the liquid phase cannot be the equilibrium phase at $\mu>\mu_c$. Clearly, one must determine when the liquid solution is physically relevant; when it is not,
one should find other solutions \cite{BiroliMezard01}.

\subsection{Linear stability limit of the liquid}
\label{subsec:3:4}

The Bethe-Peierls approximation fails when the bulk properties of the Cayley tree become sensitive to the boundary conditions.
This ``instability'' may show up via a loss of stability of the fixed point
equations as given by a simple linear analysis. Indeed, starting with fields identically and independently distributed on the leaves, the field
distribution $\mathcal{P}_{g+1}(h)$ at ``generation'' $g+1$ is related to that at
generation $g$ by
\begin{equation}
\label{eq:rec}
\mathcal{P}_{g+1}(h_i)=\int \prod_{j=1}^k dh_j\mathcal{P}_g(h_j) \delta \left(h_i-\hat h(h_1,\dots,h_k)\right).
\end{equation}
Close to a liquid solution we have to first order
\begin{equation}
\label{eq:jacob}
\langle \delta h \rangle_{g+1}=k\frac{\partial \hat h}{\partial h_1}\bigg|_{\rm liq}\langle \delta h\rangle_g
\end{equation}
where $\delta h\equiv h-h_{\rm liq}$ and $\langle .\rangle_g$ refers to the average using the distribution $\mathcal{P}_g$.
Since $h$ is a two-component vector, $\partial \hat h/\partial h_1$ is actually a $2\times 2$ Jacobian matrix. If $\lambda_1$ denotes the
eigenvalue of largest modulus of that matrix, the stability criterion simply reads
\begin{equation}
k|\lambda_1|\leq 1.
\end{equation}
 When $k|\lambda_1|> 1$, the liquid solution is unstable to perturbations homogeneous within a generation; we shall refer to it as the {\it modulation instability} because it is a transition to a regime with successive (homogeneous) generations carrying different fields.

An alternative point of view consists in studying response functions. At finite $\mu$, the response to a perturbation is related to correlations through the fluctuation-dissipation theorem, and an instability can be detected by means of a diverging susceptibility. Thus, we recover the previous result by computing the
{\it linear susceptibility} in the liquid phase, looking for the point
where it becomes infinite. The linear susceptibility is defined in terms of
connected correlation functions $\langle n_in_j\rangle_c$ via
\begin{equation}
\chi_1(\mu)\equiv\frac{d\rho}{d\mu}=\frac{1}{N}\sum_{i,j}\langle n_in_j\rangle_c.
\end{equation}
Making use of the homogeneity of the liquid solution, it can be rewritten
\begin{equation}
\chi_1(\mu)=\rho (1-\rho)+\sum_{r=1}^\infty (k+1)k^{r-1}\langle n_0n_r\rangle_c
\end{equation}
where $n_0$ and $n_r$ are taken at distance $r$ in the tree. The series converges
provided that
\begin{equation}
\label{eq:stabcrit}
\ln k+\lim_{r\to\infty}\frac{\ln\langle n_0n_r\rangle_c}{r}<0.
\end{equation}
To evaluate $\langle n_0n_r\rangle_c$, we invoke the
fluctuation-dissipation relation
\begin{equation}
\langle n_0n_r\rangle_c=\frac{\partial \langle n_r\rangle}{\partial h_0^{(c)}}
\end{equation}
where $h_0^{(c)}$ denotes the external field conjugate to $n_0$. Since
$h_0^{(c)}$ is a function of (the components of) $h_0$ only, we can use
the chain rule
\begin{equation}
\label{eq:chain}
\frac{\partial\langle n_r\rangle}{\partial h_0^{(c)}}=\frac{\partial\langle n_r\rangle}{\partial h_{r-1}}\left(\prod_{l=1}^{r-1}\frac{\partial h_l}{\partial h_{l-1}}\right)\frac{\partial h_0}{\partial h_0^{(c)}}
\end{equation}
where we introduced $h_l\equiv \hat h(h_{l-1},h_{\rm liq}\dots ,h_{\rm liq})$  as intermediate fields. In the liquid phase, all these fields are equal and
the previous equation factorizes, leading again to $k|\lambda_1|\leq
1$. For instance for the BM, the modulation instability shows up at $\mu_{\rm mod}=4\ln 2\simeq2.77$, well before the entropy becomes negative at $\mu_{s=0}\simeq 7.40$ as shown in Fig.~\ref{fig:liquid}.

\subsection{Crystal phase}
\label{subsec:3:5}

We can ask whether it is possible to choose boundary conditions such that the interior of the Cayley tree has a periodic structure (for a general tree, the boundary conditions will vary from leaf to leaf).
In the $\mu\to\infty$ limit, we expect a crystalline phase to exist whose structure can be easily displayed by starting at the center of the Cayley tree, filling the sites with particles whenever possible. In this case, for the BM, the field $h=(a,b)$ takes three different values $h_e$, $h_u$ and $h_s$ such that
\begin{equation}
\label{eq:crystal}
\begin{split}
h_e=\hat h(h_s,h_s),\\
h_u=\hat h(h_e,h_e),\\
h_s=\hat h(h_e,h_u),
\end{split}
\end{equation}
with $h_e=(0,0)$, $h_u=(1,1)$ and $h_s=(1,0)$. The integer values 0 and 1 of the components of these fields reflect the fact that a given site is certainly empty, unsaturated or saturated. For finite $\mu$, fluctuations are present but we can still look for a fixed point of Eq.~(\ref{eq:crystal}), with  $h_e, h_u, h_s \in \mathbb{R}$.

Such a solution, distinct from the liquid one, is found to exist and to
be stable for $\mu\geq \mu_{ms}$ with $\mu_{ms}\simeq 2.89$ ($ms$ for {\it melting spinodal}). 
Next we want to evaluate the grand potential of this crystalline
solution in the bulk. To do so, we first consider the $\mu=\infty$ limit and estimate the density of empty and occupied sites, $\rho_0=2/5$
and $\rho_1=3/5$, and also the proportion of edges connected to one
and two particles, $\pi_1=4/5$ and $\pi_2=1/5$. Then the crystalline potential can be written as $\omega_{\rm cryst}=\omega_{\rm site}-3/2\ \omega_{\rm edge}$, with
\begin{equation}
\label{eq:potcrystal}
\begin{split}
\omega_{\rm site}&=\rho_0\Delta\hat\Omega_{\rm site}(h_s,h_s,h_s)+\rho_1\Delta\hat\Omega_{\rm site}(h_u,h_e,h_e),\\
\omega_{\rm edge}&=\pi_1\Delta\hat\Omega_{\rm edge}(h_e,h_s)+\pi_2\Delta\hat\Omega_{\rm edge}(h_u,h_u).
\end{split}
\end{equation}
Now for $\mu_{ms}<\mu<\infty$, the structure of the crystal is preserved, with empty and occupied sites being replaced by {\it most probably} empty or occupied site. Therefore we resort to Eq.~(\ref{eq:potcrystal}), using the adequate values of the fields $h_e$, $h_u$, $h_u$, and we keep the same factors $\rho_{0,1}$ and $\pi_{0,1}$.
This leads to a melting transition (where $\omega_{\rm liq}=\omega_{\rm cryst}$) at $\mu_m\simeq3.24$.

Comparing $\mu_{ms}\simeq 2.89$ with the location of the liquid's instability
$\mu_{\rm mod}\simeq 2.77$, we have an interval $\mu_{\rm mod}<\mu<\mu_{ms}$ where no homogeneous nor periodic solution seems to exist.
This is to be contrasted with the phase diagram of the hard sphere
model studied by Runnels \cite{Runnels67} where the presence of a
particle on a site forbids the occupation of any of its neighboring
sites. That model has been reconsidered recently in two dimensions as
a combinatorial problem of counting binary matrices with no two
adjacent 1's when $\mu=0$ \cite{Baxter99} and on random graphs as an
optimization problem called {\it vertex cover problem} when
$\mu=\infty$ \cite{WeigtHartmann01}; from the point of view of our
lattice glasses, these models correspond to the $\ell=0$ constraint. In
this case the modulation instability coincides exactly with the melting
transition (and with the spinodal point), $\mu_m=\mu_{ms}=\mu_{\rm
mod}$. This is due to the special structure of the crystal, organized
in alternate shells of empty and occupied sites, and therefore
described by a cyclic solution of the liquid equation,
$h_1=\hat h(h_0,\dots,h_0)$ and $h_0=\hat h(h_1,\dots, h_1)$ with
homogeneous shells (i.e., the arguments of $\hat h$ are all
identical). For $\ell=1$ no such cyclic solution was found and
homogeneous boundary conditions on the leaves yield an {\it aperiodic}
behavior of the recursion $h_{j+1}=\hat h(h_j,\dots,h_j)$.
This feature is very specific to the unphysical nature 
of pure Cayley trees: it does not survive in the random 
regular graphs which we use below.
Therefore we have not pushed its study any further.

\section{Models on random regular graphs}
\label{sec:4}

\subsection{From Cayley trees to random regular graphs}
\label{subsec:4:1}

When the Bethe-Peierls approximation no longer holds on a Cayley tree, the
sensitivity to boundary conditions does not allow one to define a
thermodynamic limit and therefore may lead to unphysical results. Since that is due to the presence of a finite fraction of the sites on the
surface, one way to get rid of this problem is to define Bethe lattice models as models on {\it random regular graphs}. These are simple graphs with
fixed connectivity $k+1$, {\it simple} meaning that they have no
trivial loops (joining a site to itself) and no multi-edges (no two edges
join the same sites). Here we will use the cavity method to obtain results for a
typical random graph (chosen uniformly in the set of all random 
regular graphs with a fixed connectivity $k+1$).

When the number $N$ of sites is large, typical random regular graphs look locally like trees, having only long loops of order $\ln N$. Therefore the recursive equations still (locally) hold, making these lattices analytically tractable. The large loops implement an analog of generic boundary conditions and the resulting frustration forbids crystalline orderings. One then expects the system to possess an equilibrium glass phase in the high $\mu$ region.

In the low $\mu$ phase, the liquid solution is recovered. The
corresponding Bethe-Peierls approximation is called the {\it factorized
  replica symmetric approximation} in the context of the cavity method
where the vocabulary is inherited from the  treatment of spin
glasses based on the replica trick. However the glassy high $\mu$
regime will be characterized by the existence of many solutions of the
local equations and we will have to resort to the replica symmetry
breaking (rsb) formalism to correctly take into account the
specific organization of these solutions \cite{MezardParisi01}. 

\subsection{Entropy crisis}
\label{subsec:4:2}

 Reconsidering the stability of the liquid with the modulation instability now being excluded, we can look for a ``spin-glass'' instability, to borrow the terminology from magnetic systems where the modulation instability is referred to as the ``ferromagnetic instability''. This new instability manifests itself as a divergence of the {\it non-linear susceptibility}, which is defined as
\begin{equation}
\chi_2(\mu)\equiv\frac{1}{N}\sum_{i,j}\langle n_in_j\rangle^2_c.
\end{equation}
In the formalism of Sect.~\ref{subsec:3:4}, the instability appears as a widening of the variance $\langle (\delta h)^2\rangle_g$ under the recursion of Eq.~(\ref{eq:rec}). Both approaches are equivalent and lead to a stability criterion
\begin{equation}
k|\lambda_1|^2\leq 1.
\end{equation}
The eigenvalue $\lambda_1$ is the same as for the linear susceptibility since the transfer matrix is just the square of the Jacobian matrix we had in Eq.~(\ref{eq:jacob}). Note that this condition is always weaker than that for the modulation instability, $k|\lambda_1|\leq 1$. However it is the relevant one in the case of random graphs where homogeneous perturbations are generically incompatible with frustration.
If the liquid is locally stable for all $\mu$,
a continuous phase transition is excluded. For $\ell =1$ this happens for $k=2,3$ because $\sqrt{k}|\lambda_1(\mu)|< 1$ for all $\mu$ with only asymptotically $\sqrt{k}|\lambda_1(\mu)|\to 1$ as $\mu\to \infty$. In the general case, as soon as $\mu_g>\mu_{s=0}$, where $\mu_g$ ($g$ for {\it glass}) is defined by
\begin{equation}
\sqrt{k}|\lambda_1(\mu=\mu_g)|=1,
\end{equation}
the resolution of the entropy crisis requires a phase transition before the spin-glass local instability is reached, and we conclude that a {\it discontinuous} phase transition must take place at $\mu_c\leq\mu_{s=0}$. In that case we expect
a behavior similar to that of infinite-connectivity models solved
within a one-step replica symmetry breaking (1-rsb) Ansatz, like e.g. the
$p$-spin models ($p>2$). When  $\mu_g<\mu_{s=0}$, as is found for $\ell=1$ and $k\geq 4$, we can either have a continuous transition at $\mu_g$ or a discontinuous transition at $\mu_c<\mu_g$; a study of the {\it local} stability of the liquid solution says nothing about which case arises.

\subsection{Cavity equations}
\label{subsec:4:3}

The solution by the cavity method predicts results for quantities averaged over all random regular graphs with size $N\to\infty$, but the problem is more clearly stated on a given finite regular graph. Indeed, we want to solve self-consistently a
set of $(k+1)N$ coupled equations for the cavity fields
\begin{equation}
\label{eq:TAP}
\begin{split}
&h_{i\to j_0}=\hat h(h_{j_1\to i},\dots,h_{j_k\to i})\\
&\forall i \ \ \ \forall \{j_0,j_1,\dots,j_k\}\in \mathcal{N}(i)
\end{split}
\end{equation}
where $\mathcal{N}(i)$ denotes the set of $k+1$ neighbors of $i$. Thus for each $i\in \{1,\dots, N\}$ we have $k+1$ equations
corresponding to the different
choices of $j_0\in\mathcal{N}(i)$. The notation $h_{i\to j}$ refers
to the local field on $i$ when the edge to the site $j$ is removed, and it is called a {\it cavity field}; for a Cayley tree, it corresponds to the field on the root $i$ of a rooted tree obtained when
 the edge $ij$ has been removed. These local equations are known as the TAP
equations in statistical physics \cite{MezardParisi87b} and as the belief propagation equations in computer science \cite{KschischangFrey01}. The contact between
both points of view has recently lead to establish that these equations must always
have at least one fixed point corresponding to the minimum of a
correctly defined Bethe-Peierls approximated free energy \cite{YedidiaFreeman02}. Their solutions should correspond to the fuzzy concept of {\it states} ubiquitous in the spin-glass literature.

A message passing algorithm can be used to try to solve these
equations on a given graph of large but finite size $N$. First on each
oriented edge we associate local fields $h_{i\to j}$ randomly
initialized. Then we proceed iteratively: at each step, all the
oriented edges are successively chosen in a random order, and the
field on the chosen oriented edge is updated by taking into account
the values of its neighbors as prescribed by Eq.~(\ref{eq:TAP}). The
iteration is stopped when a sweep of all the oriented edges results in
no change; in such a case we lie at a fixed point of the local
equations. However the algorithm may also not converge, in which case
no conclusion can be drawn. In practice, we find a rapid convergence
toward the liquid solution for $\mu<\mu_{bp}$ and then a failure to
converge for $\mu>\mu_{bp}$. The critical $\mu_{bp}$ depends slightly
on the graph but for large $N$ and large $k$, it is given by the glass
instability, i.e., $\mu_{bp}\to\mu_g$ as $k\to\infty$. In some case
(for not too big graphs), we could find a convergence towards a
non-liquid distribution, suggesting that the high $\mu$ region
corresponds in fact to a glassy phase. In order to deal with this
phase where the Bethe-Peierls approximation breaks down and simple
message passing algorithms fail, we will now introduce the cavity
method which provides both an alternative approximation for infinite
graphs and insights into elaborating more efficient algorithms for
finite graphs \cite{MezardParisi02}. In Appendix D we present an
alternative approach based on the replica method.

\subsection{One-step replica symmetry breaking cavity method}
\label{subsec:4:4}

The local equations Eq.~(\ref{eq:TAP}) can be written for arbitrary graphs but they provide exact marginal probability distribution (and thus exact particle densities) only for trees. In addition, they are in general intractable. However, they are particularly suited for very large random graphs, where due to the local tree-like structure, they are expected to provide good approximations of the marginals and where additional hypotheses allow for an analytical treatment. To do so, we
do not try to find one solution but instead turn to a statistical
treatment of sets of solutions.

Being interested in the case where many solutions exist, we fix a $\mu$ for which 
this is supposed to happen. We make the further hypothesis that exponentially many (in $N$) solutions exist. 
More precisely, we assume that the number
$\mathcal{N}_N(\omega)$ of solutions with a given potential density
$\omega$ on graphs of size $N$ is given by
\begin{equation}
\label{eq:complexity}
\mathcal{N}_N(\omega)\sim\exp[N\Sigma (\omega)]
\end{equation}
where $\Sigma (\omega)\geq 0$ is called the {\it configurational entropy} (or {\it complexity}) and is supposed to be an increasing and concave function of the grand potential $\omega$. This is a
strong hypothesis which is justified by its self-consistency and by
its consequences (in particular it matches with the output of replica
theory calculations, cf. Appendix~D).

Starting with a graph $\mathcal{G}$, we pick a site $i$ and one of its neighbors $j_0\in\mathcal{N}(i)$, and define the graph $\mathcal{G}_{i\to j_0}$ as the connected graph containing $i$ obtained by removing the edge $ij_0$ from $\mathcal{G}$. If $\mathcal{G}$ is a Cayley tree, $\mathcal{G}_{i\to j_0}$ is nothing but a rooted tree, the appropriate structure to write down recursive relations. We introduce $R_{i\to j_0}(h,\omega)$, the joint probability density, when a solution $\alpha$ of Eq.~(\ref{eq:TAP}) defined on $\mathcal{G}_{i\to j_0}$ is chosen randomly, that the cavity fields $h^{(\alpha)}_{i\to j_0}$ on $i$ take the value $h$ and the grand potential density $\omega^{(\alpha)}_{i\to j_0}$ the value $\omega$,
\begin{equation}
R_{i\to j_0}(h,\omega)=\frac{1}{N_{\rm sol}}\sum_\alpha\delta \left(h-h^{(\alpha)}_{i\to j_0}\right)\delta\left(\omega-\omega^{(\alpha)}_{i\to j_0}\right).
\end{equation} 

The next crucial hypothesis is based on the locally tree-like structure of large random graphs. Indeed, if the graph $\mathcal{G}_{i\to j_0}$ is a rooted tree, the sub-rooted trees $\mathcal{G}_{j_1\to i},\dots, \mathcal{G}_{j_k\to i}$ with $\{j_1,\dots,j_k\}\in \mathcal{N}(i)$ are disjoint and the cavity fields $h_{j_1\to i},\dots, h_{j_k\to i}$ are therefore solutions of uncoupled local equations. In such a case the joint probability distribution factorizes over the independent variables
\begin{equation}
\begin{split}
&R(h_{j_1\to i},\omega_{j_1\to i};\dots; h_{j_k\to i},\omega_{j_k\to i})\\
&= R_{j_1\to i}(h_{j_1\to i},\omega_{j_1\to i})\dots R_{j_k\to i},\left(h_{j_k\to i},\omega_{j_k\to i}\right).
\end{split}
\end{equation}

In addition, a simple relation between the distribution $R_{i\to j_0}$ on a rooted tree $\mathcal{G}_{i\to j_0}$ and the $R_{j_1\to i}$,\dots,$R_{j_k\to i}$ on its sub-rooted trees $\mathcal{G}_{j_i\to i}$,\dots,$\mathcal{G}_{j_k\to i}$ can be written:
\begin{equation}
\label{eq:recforR}
\begin{split}
&R_{i\to j_0}(h,\omega)=\int\prod_{r=1}^kdh_{j_r}d\omega_{j_r}R_{j_r\to i}(h_{j_r},\omega_{j_r})\\
&\delta\left(h-\hat h(\{h_{j_r}\})\right)
\delta\left[\omega-\frac{1}{N}\left(\sum_{r=1}^kN_{j_r}\omega_{j_r}+\Delta\hat\Omega_{\rm iter}(\{h_{j_r}\})\right)\right]
\end{split}
\end{equation}
where the $N_{j_r}$ are the sizes of the sub-trees and $N=\sum_rN_{j_r}+1$ is the size of $\mathcal{G}_{i\to j_0}$. A further simplification takes place on regular trees where, due to the absence of local quenched disorder, the equations for $h_{j_1\to i}$,\dots, $h_{j_k\to i}$ are identical. In such a case, $R_{i\to j_0}$ is in fact independent of the oriented edge $i\to j_0$. For a large random regular graph, we assume that the same properties hold. Note that the last hypothesis, yielding $R_{i\to j_0}=R$ is called the {\it factorization} approximation and could be relaxed by working with a distribution $\mathcal{R}[R]$ of the $R$ over the various edges \cite{MezardParisi01}. However the factorization approximation should be exact on random regular graph for systems without disorder provided that no spontaneous breaking of ``translational'' invariance occurs, and we will not consider this extension here. 

We now want to write the 1-rsb cavity equation, which is a self-consistent equation for the distribution $P^{(\omega)}$ of local fields $h$ at a {\it fixed} grand potential $\omega$. $P^{(\omega)}$ is obviously proportional to $R$, $P^{(\omega)}(h)\propto R(h,\omega)$, and since the distribution of the $\omega$ is given by $\int dh R(h,\omega)=C\exp[N\Sigma(\omega)]$, we have the relation
\begin{equation}
\label{eq:Rpropto}
R(h,\omega)=C e^{N\Sigma(\omega)}P^{(\omega)}(h)
\end{equation}
with $C$ a proportionality constant independent of both $h$ and $\omega$.
Next we fix a grand potential density $\omega_0$ and consider only potentials $\omega$ close to $\omega_0$, noted $\omega\in \mathcal{V}_{\omega_0}$, such that we can linearize the complexity
\begin{equation}
\Sigma(\omega)\simeq \Sigma(\omega_0)+m\mu(\omega-\omega_0)
\end{equation}
with
\begin{equation}
\label{eq:defm}
m(\omega_0)\equiv\frac{1}{\mu}\frac{d\Sigma}{d\omega}(\omega_0).
\end{equation}
Given the concavity of the complexity $\Sigma$, fixing $\omega_0$ is equivalent to fixing $m$ and thus $\mathcal{V}_{\omega_0}$ can be rewritten as $\mathcal{V}_m$. Plugging relation~(\ref{eq:Rpropto}) into Eq.~(\ref{eq:recforR}), we find that the distribution defined by
\begin{equation}
P^{(m)}(h)\propto\int_{\omega\in\mathcal{V}_m}d\omega P^{(\omega)}(h)
\end{equation}
satisfies a simple self-consistent equation
\begin{equation}
\label{eq:cavity1-rsb}
\begin{split}
&P^{(m)}(h)\propto\\
&\int\prod_{j=1}^kdh_j P^{(m)}(h_j)\delta\left(h-\hat h(\{h_j\})\right)e^{-m\mu\Delta\hat\Omega_{\rm iter}(\{h_j\})}.
\end{split}
\end{equation}
Eq.~(\ref{eq:cavity1-rsb}) is called the {\it factorized 1-rsb cavity equation}. We will drop the explicit reference to the parameter $m$ in the ensuing discussion, but it should be kept in mind that the 1-rsb cavity field distribution $P(h)$ is $m$-dependent.

As we have seen, it is convenient to fix $m$ instead of $\omega$; going from $\omega$ to $m$ actually amount to performing a Legendre transformation. The complexity $\Sigma(\omega)$ is recovered by Legendre transforming the function $\Phi(m)$ defined as
\begin{equation}
\label{eq:defphi}
m\Phi(m)=m\omega-\frac{1}{\mu}\Sigma(\omega)
\end{equation}
via the relation
\begin{equation}
\label{eq:Sigma}
\frac{1}{\mu}\Sigma(\omega)=m^2\partial_m\Phi(m).
\end{equation}

Eq.~(\ref{eq:defphi}) is similar to the definition of the entropy through
\begin{equation}
\mu\omega(\mu)=-\mu\rho-s(\mu)
\end{equation}
[cf. Eq.~(\ref{eq:defentropy})]. Indeed $\Phi(m)$ is for the
states the analog of the grand potential $\omega(\mu)$ for the
configurations, the parameter $m$ sampling the states as the chemical
potential $\mu$ samples configurations, and the complexity
$\Sigma(\omega)$ counting states as the entropy $s(\mu)$ counts
configurations \cite{Monasson95}. The quantity corresponding to the grand partition function is
\begin{equation}
\label{eq:Xi(m,mu)}
\begin{split}
\Xi(m)&=\sum_\alpha \exp[-Nm\mu \omega^{(\alpha)}(\mu)]\\
&=\int d\omega \exp \left(N[\Sigma(\omega)-m\mu\omega]\right)\\
&=\exp [-Nm\mu\Phi(m)].
\end{split}
\end{equation}
Following the same analogy, $\Phi(m)$ can be computed similarly to a grand potential. To do so, we first need to generalize to random regular graphs the construction that led us to the Bethe-Peierls approximation of the grand potential, Eq.~(\ref{eq:betheomega}). Likewise, we want to write
\begin{equation}
\label{eq:phi=}
\Phi(m)=\Delta\Phi_{\rm site}(m)-\frac{k+1}{2}\Delta\Phi_{\rm edge}(m)
\end{equation}
with $\Delta\Phi_{\rm site}(m)$ the contribution from a site addition, and $\Delta\Phi_{\rm edge}(m)$ from an edge addition. The way two sites can be added is even simpler than for Bethe lattices here because we do not care about introducing loops: take a random regular graph of size $N$ and connectivity $k+1$, pick $(k+1)$ edges and remove them, leading to $2(k+1)$ amputated sites with $k$ neighbors instead of $(k+1)$. Then add two new sites and connect each one to $(k+1)$ of the amputated sites, leading to a new random regular graph of size $N+2$ and same connectivity $(k+1)$. Thus the contribution to $\Phi$ when going from $N$ to $N+2$ sites is equivalent to the contribution from two site additions plus $(k+1)$ edge deletions [i.e., minus $(k+1)$ edge additions], as expressed by Eq.~(\ref{eq:phi=}).

To see how this is related to the grand potential shifts $\Delta\Omega_{\rm site}$ and $\Delta\Omega_{\rm edge}$, we rewrite Eq.~(\ref{eq:Rpropto}) introducing the definition Eq.~(\ref{eq:defm}) of the parameter $m$ as
\begin{equation}
R(h,\omega)=e^{m\mu N\left(\omega-\omega^{(0)}\right)}P(h)
\end{equation}
where $\omega^{(0)}=\omega^{(0)}(m)$ enforces the normalization. The function
\begin{equation}
\rho_0(\Omega)\equiv e^{m\mu\left(\Omega-\Omega^{(0)}\right)}
\end{equation}
gives the distribution of grand potential $\Omega=N\omega$ on a rooted graph of size $N$ with one site having only $k$ neighbors. Now we take $k+1$ such rooted graphs and determine the distribution $\rho_1(\Omega)$ when one site is added. It is given by
\begin{equation}
\begin{split}
\rho_1(\Omega)&=\int\prod_{j=1}^{k+1}dh_jd\Omega_jR(h_j,\Omega_j)\delta \left(\Omega-\sum_{j=1}^{k+1}\Omega_j-\Delta\hat\Omega_{\rm site}(\{h_j\})\right)\\
&=e^{m\mu \left(\Omega-\sum_{j=1}^{k+1}\Omega^{(0)}_j\right)}\int\prod_{j=1}^{k+1}dh_jP(h_j)e^{-m\mu\Delta\hat\Omega_{\rm site}(\{h_j\})}\\
&\equiv e^{m\mu \left(\Omega-\Omega^{(1)}\right)}
\end{split}
\end{equation}
with $\Omega^{(1)}=\sum_{j=1}^{k+1}\Omega^{(0)}_j+\Delta\Phi_{\rm site}$, while the mean shift $\Delta\Phi_{\rm site}$ due to a site addition is
\begin{equation}
\Delta\Phi_{\rm site}(m)=-\frac{1}{m\mu}\ln\left[\int\prod_{j=1}^{k+1}dh_jP(h_j)e^{-m\mu\Delta\hat\Omega_{\rm site}(\{h_j\})}\right].
\end{equation}
We compute similarly the contribution from edge addition,
\begin{equation}
\begin{split}
\rho_2(\Omega)&=\int \prod_{j=1}^2dh_jd\Omega_jR(h_j,\Omega_j)\delta \left(\Omega-\sum_{j=1}^2\Omega_j-\Delta\hat\Omega_{\rm edge}(h_1,h_2)\right)\\
&=e^{m\mu \left(\Omega-\Omega^{(0)}_1-\Omega^{(0)}_2\right)}\int\prod_{j=1}^2 dh_jP(h_j)e^{-m\mu\Delta\hat\Omega_{\rm edge}(h_1,h_2)}\\
&\equiv e^{m\mu \left(\Omega-\Omega^{(2)}\right)}
\end{split}
\end{equation}
with $\Omega^{(2)}=\Omega^{(0)}_1+\Omega^{(0)}_2+\Delta\Phi_{\rm edge}$, while the mean shift $\Delta\Phi_{\rm edge}$ due to an edge addition is
\begin{equation}
\Delta\Phi_{\rm edge}(m)=-\frac{1}{m\mu}\ln\left[\int \prod_{j=1}^2dh_jP(h_j)e^{-m\mu\Delta\hat\Omega_{\rm edge}(h_1,h_2)}\right].
\end{equation}

The replica symmetric description of the liquid phase is recovered by
taking $P(h)=\delta(h-h_{\rm liq})$. When many
solutions coexist, by varying $m=m(\omega_0)$ at fixed $\mu$, we describe
states characterized by different values of $\omega_0$ and the
question is which $m$ must be selected to describe the {\it equilibrium} (glassy) thermodynamics. 
 Following Eq.~(\ref{eq:Xi(m,mu)}), the  grand partition function is
\begin{equation}
\Xi(\mu)=\int d\omega \exp \left(N[\Sigma(\omega)-\mu\omega ]\right)
\end{equation}
and the saddle point method for $N\to\infty$ indicates that the grand potential $\omega$ of the dominating states is such that $\mu=\partial_\omega\Sigma(\omega)$. But of course, this saddle equation is relevant only if its solution $\omega_s$ lies inside the interval range $]\omega_{\rm min},\omega_{\rm max}[$ where the integral is performed, which corresponds to the range where $\Sigma(\omega)$ is strictly positive. Generically, for $\mu>\mu_s$, it is found that $\omega_s<\omega_{\rm min}$, and we must then take instead $\omega_s=\omega_{\rm min}$. 
A kind of replica trick intervenes here to balance the complexity contribution thanks to the parameter $m$, in such a way that $\omega_s$ is always given by a saddle equation.
So, in the glassy phase $\mu>\mu_s$ where the rsb formalism becomes necessary, we want the equilibrium grand potential to be given by $\omega_s=\omega_{\rm min}(\mu)\equiv\Phi(m_s)$. Since the complexity curve is expected to be continuous at $\omega_{\rm min}$, we can alternatively ask for the condition $\Sigma(\omega_s)=0$. From Eq.~(\ref{eq:Sigma}), we see that it corresponds to extremizing $\Phi (m)$, i.e.,
\begin{equation}
\partial_m\Phi(m=m_s)\equiv 0
\end{equation}
which is precisely the criterion provided by the replica method for selecting the breaking point parameter $m$.

Note that other values of $m$ also carry physical information. Lower
values of $m$ ($m<m_s$) describe metastable states for which
$\omega>\omega_s$ (it is the analog of a non-zero temperature giving
access to excited configurations); of particular interest is the
value $\omega_d$ associated to the maximum complexity $\Sigma
(\omega_d)\equiv\max_\omega\Sigma(\omega)$ since its describes the
most numerous states.
We expect that this is the portion of phase space 
where the system will get almost trapped at long times after a ``quench''
from the low density liquid phase.

Higher values of $m$ ($m>m_s$) are usually termed as unphysical, but in fact they describe properties of systems associated with untypical graphs. In particular, it can be shown that $m=1$ always gives back the liquid solution. In fact, for $m>m_s$, the cavity method leads to a negative complexity, which seems to be in contradiction with its initial definition Eq.~(\ref{eq:defm}). The point is that in the cavity method, the graph is not specified and $e^{N\Sigma(\omega)}$ is the number of states with grand potential $\omega$ after averaging over different graph realizations. This fact has no consequence whenever the average corresponds to the typical case, which is expected as soon as $\Sigma(\omega)>0$ ; indeed some graph realizations may behave very differently from typical realizations, but their contribution to the averaged quantities is negligible. However, an exception is worth mentioning: if the quantity we average is typically strictly zero, but happen to be positive for exponentially rare realizations, it leads to a small but non zero (i.e. non typical) average. We find such a behavior here, where some untypical graphs allow for $\omega_s$ lower than the typical value $\omega_s^{(\rm typ)}$, leading to a $\Sigma\left(\omega<\omega_s^{(\rm typ)}\right)<0$ even if the complexity on a given graph is intrinsically a positive quantity.

An analog phenomenon happens for instance in the random energy model, where averaging over disorder leads to a negative entropy associated with energies lower than the typical ground state, while for a given realization of the disorder the entropy is necessarily positive. Here the role of the quenched disorder is taken by the topological disorder from the various realizations of random regular graphs. From this point of view, a random graph with no frustrating loop is an example of untypical graph which has a crystalline ground state, as opposed to the glassy ground states of typical random regular graphs.

\subsection{Observables}
\label{subsec:4:6}

To complete our overview of the cavity method, we now show on the example of the particle density how physical observables can be computed; Sect.~\ref{subsec:4:10} will provide an other example with the computation of susceptibilities. We begin by considering rooted trees where the particle density on the root $i$ is simply given by
\begin{equation}
\langle n_i \rangle_{\rm rooted\ tree}=\frac{\Xi^{(s)}_i+\Xi^{(u)}_i}{\Xi^{(s)}_i+\Xi^{(u)}_i+\Xi^{(e)}_i}=1-e^{-\mu a_i}.
\end{equation}
Now for a Cayley tree, we need to take into account $k+1$ neighbors instead of $k$,
\begin{equation}
\langle n_i \rangle_{\rm Cayley\ tree}=1-e^{-\mu A_i}.
\end{equation}
where the total local field $A$ is computed similarly to the cavity field $a$ but substituting $k$ by $k+1$, i.e.,
\begin{equation}
\begin{split}
A_i&=\hat A(a_1,b_1,\dots,a_{k+1},b_{k+1})\\
&=\frac{1}{\mu}\ln \left[1+e^{\mu (1-\sum_{j=1}^{k+1}a_j)}\left(1+\sum_{j=1}^{k+1}(e^{\mu b_j}-1)\right)\right].
\end{split}
\end{equation}
The total field $B_i$ would be defined similarly; note that with our notations, we simply have $\hat A=-\Delta\hat\Omega_{\rm site}$. On a random graph in the 1-rsb phase, we need to include the reweighting associated with the addition of the site $i$:
\begin{equation}
\begin{split}
&\rho(\mu,m)\\
&=\frac{\int\prod_{j=1}^{k+1} dh_jP(h_j)(1-e^{-\mu \hat A(\{h_j\})})e^{-m\mu\Delta\hat\Omega_{\rm site}(\{h_j\})}}{\int\prod_{j=1}^{k+1} dh_jP(h_j)e^{-m\mu\Delta\hat\Omega_{\rm site}(\{h_j\})}}
\end{split}
\end{equation}
and the equilibrium value is given by $\rho(\mu,m_s)$.
The same lines can be followed to compute any other observables. 

\subsection{Order parameters}

Comparing with the replica method \cite{MezardParisi87b}, the cavity approach focuses on local fields instead of overlaps between states. 
 In lattice glass models, the latter can be defined as
\begin{equation}
q_{\alpha\beta\gamma\dots}\equiv \frac{1}{N}\sum_{i=1}^N\langle n_i\rangle_\alpha\langle n_i\rangle_\beta\langle n_i\rangle_\gamma\dots
\end{equation}
with the indices $\alpha$, $\beta$, $\gamma$, \dots denoting states randomly chosen according to their Boltzmann weights.

Overlaps are particularly useful in infinite connectivity systems where, due to the central limit theorem, the first two moments $q_\alpha$ and $q_{\alpha\beta}$ are sufficient to encode the Gaussian distribution of the cavity fields. In contrast, for finite connectivity systems, an infinite number of overlaps must be kept and working directly with the local field distribution is simpler. However, the two choices of order parameters, local cavity fields or global overlaps, provide complete and equivalent descriptions; in particular overlaps can be easily recovered from the knowledge of the cavity field distribution.

At the replica symmetric level, all indices $\alpha$, $\beta$, $\gamma$, \dots are equivalent, so we only need to distinguish overlaps according to the number $r$ of states they involve, and we have
\begin{equation}
q^{(\rm rs)}_r\equiv \overline{\langle n_i\rangle^r}=\int dAdB\mathcal{P}_{rs}(A,B)\left(1-e^{-\mu A}\right)^r
\end{equation}
with $A$ and $B$ being the total local fields of Sect.~\ref{subsec:4:6}.
Note that the replica symmetric approximation in principle already involves a functional order parameter $\mathcal{P}_{rs}$, but in our case where the factorization Ansatz is taken, the liquid distribution is trivial and we merely have $q^{(\rm rs)}_r=(1-e^{-\mu A_{\rm liq}})^r$, i.e. the order parameter is a single scalar. 

At the one-step level, we need to distinguish whether randomly chosen states are distinct or identical. Thus for overlaps involving two replicas, we can define two parameters, $q_0\equiv q_{\alpha\beta}$ ($\alpha\neq\beta$) and $q_1\equiv q_{\alpha\alpha}$, corresponding respectively in spin glasses to the Edwards-Anderson order parameter and the (squared) magnetization. Their computation amounts to calculating densities inside a state, which was the subject of the previous section. Note however that $q_0$ and $q_1$ provide only a partial description of the system and the full order parameter has here a functional structure $P(h)$. Any new level of replica symmetry breaking will require considering a more sophisticated order parameter, namely a distribution over the order parameters from the previous level. For instance, the two-step order parameter will be written as a distribution $Q[P]$ over distributions $P(h)$. Describing with this formalism a finite connectivity system with full replica symmetry breaking is therefore rather complicated, and we will limit ourself to at most two levels of replica symmetry breaking in the ensuing discussion.

\subsection{Solution via population dynamics}
\label{subsec:4:8}

Given the cavity equations (\ref{eq:cavity1-rsb}), we would like to solve them. Since they are essentially functional relations, an analytical treatment is not possible in general. An important exception however is the close-packed limit $\mu\to\infty$ ; the next section is devoted to this case. Here we consider the more general finite $\mu$ situation and use the {\it population dynamics algorithm} of Ref.~\cite{MezardParisi01} to obtain numerical results. The principle of the algorithm is elementary: the distribution
$P(h)$ is encoded in a family of $M$ fields $\{h_i\}_{i=1,\dots,M}$ such that
\begin{equation}
P(h)\simeq\frac{1}{M}\sum_{i=1}^M\delta(h-h_i),
\end{equation}
and the cavity field distribution is expressed as the fixed point of the iteration equation
\begin{equation}
\begin{split}
P_{g+1}(h)\propto&\int\prod_{j=1}^kdh_jP_g(h_j)\delta \left(h-\hat h(h_1,\dots,h_k)\right)\\
&e^{-m\mu\Delta\hat\Omega_{\rm iter}(h_1,\dots,h_k)}.
\end{split}
\end{equation}
At each step, $M$ new {\it children} are generated; each is obtained by choosing randomly $k$ {\it parents} among the population. To
take into account the reweighting, we duplicate or eliminate the
children according to their weight $e^{-m\mu\Delta\Omega_{\rm iter}}$ so that we
  keep a total population of (approximately) $M$ individuals.

However, such a recursion turns out to be unstable and we stabilize it
by means of a relaxation parameter $\epsilon\in]0,1[$. At
each step, only a fraction $\epsilon M$ of the population is regenerated. The reason for this relaxation will be made clearer when we will discuss the stability of the cavity method solution; it will be associated to the instability of the first kind discussed in Sect.~\ref{subsubsec:5:2:2}.

\subsection{Static and dynamical transitions}
\label{subsec:4:9}

At low $\mu$, the population dynamics algorithm always converges to
the liquid solution, i.e., starting from a population with an arbitrary
distribution it converges to a population of identical fields
corresponding to the fixed point $h_{\rm liq}$ of $\hat h$. When
$\mu$ is increased, a first non-trivial distribution is found at
$\mu=\mu_d$ for $m=1$. At this point many states exist, but they are
only metastable and the statics is still given by the (paramagnetic)
liquid state. This ergodicity breaking is called a dynamical phase
transition because it is where the equilibrium dynamics should display an
ergodic-non ergodic transition (see however the discussion in Sect.~\ref{sec:6}).  

A static phase transition, which is the one relevant at
equilibrium, only appears for higher chemical potential, $\mu=\mu_s$,
when the configurational entropy vanishes. At this point the grand
potential of the 1-rsb solution becomes lower than the one of the
liquid and the equilibrium phase transition takes place. 

In practice, the dynamical transition point $\mu_d$ is found by decreasing $\mu$ and looking for the $\mu$ where the 1-rsb solution disappears. For the BM, it happens at $\mu_d\simeq 6.4$ ; more generally for $\ell=1$ and other $k$ we find $\mu_d<\mu_s$, indicating that the transition is always discontinuous, as could not be directly inferred from the arguments of Sect.~\ref{subsec:4:2}. To obtain $\mu_s$, we calculate explicitly $\partial_m\phi(m,\mu)$ and look for the $\mu$ at which $\partial_m\phi(m=1,\mu)=0$; for the BM, we obtain $\mu_s\simeq 7$.

\subsection{Stability of the one-step solution}
\label{subsec:4:10}
\begin{figure}
\resizebox{0.5\textwidth}{!}{
\centering
\includegraphics{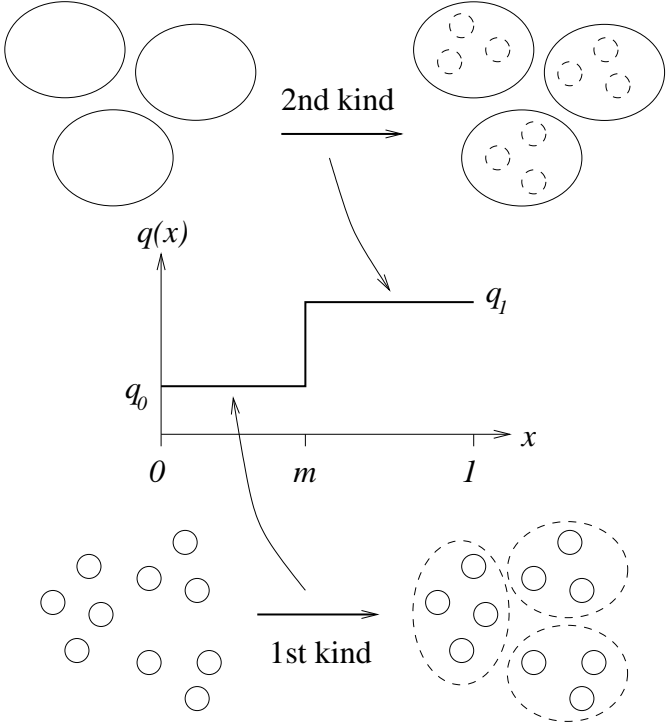}
}
\caption{Pictorial view of the two possible instabilities of the 1-rsb Ansatz. At the bottom, the states stay states but clusterize (first kind). At the top, the states become cluster of new states (second kind). If we were on a totally connected graph where an overlap function $q(x)$ can be defined, its 1-rsb shape would be affected on different parts depending on which instability is relevant; indeed its left part ($x<m$) corresponds to the inter-state overlap $q_0$ and its right part ($x>m$) to the intra-state overlap $q_1$.}
\label{fig:instabilities}
\end{figure}
To determine whether the equilibrium state is really described by a
1-rsb solution or whether further replica breakings are necessary, one
has to study the stability of the 1-rsb solution. In this section 
we set up the formalism needed to check it. In the following sections
we will analyze the stability in the close packing limit ($\mu
\rightarrow \infty $). 

In the 1-rsb phase the Gibbs measure is decomposed in a cluster
of different thermodynamic pure states \cite{MezardParisi87b}.
Thus, there are two different types of instabilities that
can show up \cite{MezardParisi87b}.
First kind: The states can aggregate into different clusters (see Fig.~\ref{fig:instabilities}). 
To study this instability one has to compute inter-state
susceptibilities:
\begin{equation}
\chi^{\rm inter}_{p}=\frac{1}{N}\sum_{i,j}\left(\overline{\langle
n_i\rangle \langle  n_j\rangle}-\overline{\langle n_i \rangle}\ \overline{\langle n_j \rangle}\right)^p.
\end{equation}
where the overline denotes an average over the states taken with their Boltzmann weights.
Second kind: Each state can fragment in different states (see
Fig.~\ref{fig:instabilities}). 
To study this instability one has to compute intra-state
susceptibilities: 
\begin{equation}
 \chi^{\rm intra}_{p}=\frac{1}{N}\sum_{i,j}\overline{\langle n_{i}n_{j}\rangle^{p}_{c}}.
\end{equation}
If any of the intra or inter-state susceptibilities diverge
then the 1-rsb glass phase is unstable (toward a 2-rsb glass phase). However, as for the liquid, the linear susceptibilities $\chi_{1}$ are
related to instabilities incompatible
with the underlying random graph structure and, hence, they are
irrelevant for our purposes. In the following we will focus on the $p=2$ case which is
the only relevant one since all the susceptibilities with $p>2$ are clearly
bounded in modulus by the $p=2$ one. 

Because of the homogeneity of the simple random graphs that we are
focusing on, the stability analysis is simplified and in particular: 
\begin{equation}
\begin{split}
\chi_2^{\rm inter}&=\sum_{r=1}^\infty
(k+1)k^{r-1}\left(\overline{\langle n_0\rangle \langle  n_r\rangle}-\rho^2\right)^2,\\
 \chi_2^{\rm intra}&=\sum_{r=1}^\infty(k+1)k^{r-1}\overline{\langle n_0 n_r\rangle^{2}_{c}}
\end{split}
\end{equation}
where $n_0$ and $n_r$ are at distance $r$ (we omitted the unimportant $r=0$ term).

We expect, as it can be proved (see below), that the correlation
functions decay exponentially at large distance
\begin{equation}
 \overline{\langle n_0 n_r\rangle_{c}^2}\sim \exp (-r/\xi_{2}),
\end{equation}
and
\begin{equation}
\left(\overline{\langle n_0\rangle \langle
n_r\rangle}-\rho^2\right)^2 \sim \exp (-r/\zeta_{2}).
\end{equation}

Due to the tree-like structure of the lattice, the resulting stability 
conditions are different from the condition $\xi<\infty$ used 
for finite dimension lattices and reads
\begin{equation}
\xi_{2}<\frac{1}{\ln k}\qquad ,\qquad \zeta_{2}<\frac{1}{\ln k}.
\end{equation} 

In the following we show how the correlation length $\xi_{p}$ can be computed.
A very similar procedure can be carried out for $\zeta_{2}$.  
In Appendix B 
we shall show explicitly how this can be done in the close-packing limit ($\mu=\infty$).

\begin{figure}
\resizebox{0.5\textwidth}{!}{
  \includegraphics{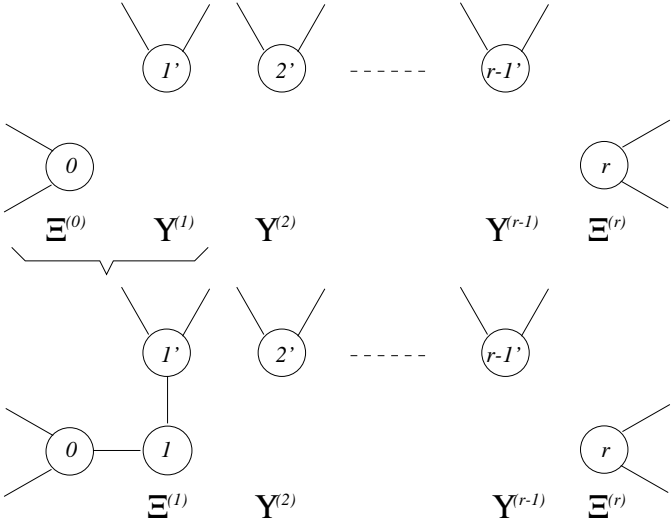}
}
\caption{Partition functions involved in the reweighting for the computation at the 1-rsb level of the two-site correlation $\overline{\langle n_on_r\rangle_{\alpha, c}^p}$, illustrated here with $k=2$. We start with the partition functions $\Xi^{(0)},\Upsilon^{(1)},\dots,\Upsilon^{(r-1)},\Xi^{(r)}$ where $\Upsilon^{(l)}$ is in general the product of the partition functions of $k-1$ rooted trees (top of the figure); for $k=2$ as illustrated, it reduce to one rooted tree with root site noted $l'$. Next we merge the $k$ rooted trees corresponding to $\Xi^{(0)}$ and $\Upsilon^{(1)}$ into a site $1$ and call the partition function of the resulting rooted tree  $\Xi^{(1)}$ (bottom). Recursively, we define similarly the other  $\Xi^{(l)}$ for $2\leq l\leq r-1$.}
\label{fig:reweightings}
\end{figure}

In order to obtain $\xi_p$ we need  to
compute the correlation functions $\overline{\langle n_0 n_r\rangle_{c}^p}$.
\begin{figure}
\resizebox{0.5\textwidth}{!}{
  \includegraphics{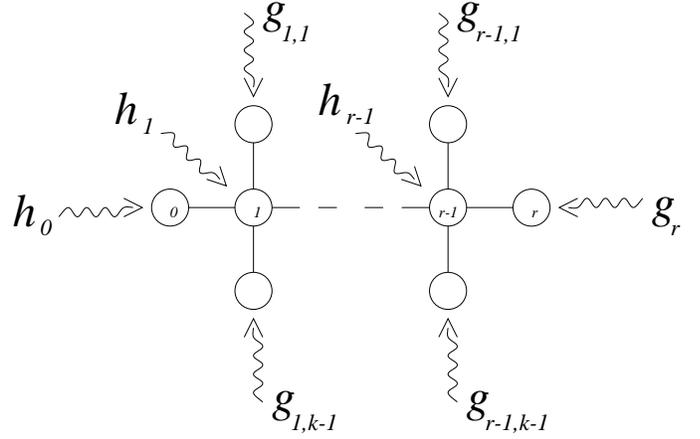}
}
\caption{Cavity diagram for computing a two-site correlation function $\langle n_0n_r\rangle_c$. As all we know is $P(h)$, the distribution of the local field on the root of a rooted tree, we build a chain of sites $l=0,\dots,r$ out of rooted trees with fields $h_0,g_{1,j},\dots,g_{r-1,j},g_r$. We proceed recursively: we first add site $l=1$, obtain a new rooted tree with root $1$ and local field $h_1=\hat h(h_0,g_{1,1},\dots,g_{1,k-1})$ so the length of the chain to compute is reduced from $r+1$ to $r$. Then we proceed further by adding site $2$, etc (see also Fig.~\ref{fig:reweightings}).}
\label{fig:susceptibiltydiagram}
\end{figure}
We write the generalization of Eq.~(\ref{eq:chain}) as
\begin{equation}
\langle n_0 n_r\rangle_{\alpha ,c}=
\left(\frac{\partial\langle n_r\rangle}{\partial h_{r-1}}\right)^t\left(\prod_{l=1}^{r-1}\partial_1 \hat h(h_{l-1};g_{l,j})\right)\frac{\partial h_0}{\partial h_0^{(c)}}
\end{equation}
where the fields are $(d-1)$-dimensional vectors ($d=3$ for our
models), so that the product involves in fact $(d-1)\times(d-1)$
dimensional Jacobian matrices $\partial_1 \hat h$ (the notation
$\partial_1 \hat h$ indicates that the derivative is taken with
respect to the first field, here $h_{l-1}$, and $()^{t}$ means transposed). Note that $\hat h(h_{l-1};g_{l,j})$ is used as a short-hand notation for $\hat h(h_{l-1},g_{l,1},\dots,g_{l,k-1})$ (see  Fig.~\ref{fig:susceptibiltydiagram}). Next we need to take into account the reweighting introduced by the addition of the sites $l=1,\dots,r-1$. With a transfer-matrix approach in mind, we write it as
\begin{equation}
\begin{split}
e^{-m\mu\Delta\Omega}&=\frac{\Xi}{\Xi^{(0)}\left(\prod_{l=1}^{r-1}\Upsilon^{(l)}\right)\Xi^{(r)}}\\
&=\prod_{l=1}^{r-1}\left(\frac{\Xi^{(l)}}{\Xi^{(l-1)}\Upsilon^{(l)}}\right)\frac{\Xi}{\Xi^{(r-1)}\Xi^{(r)}}
\end{split}
\end{equation} 
where the notation refers to  Fig.~\ref{fig:reweightings}: $\Upsilon^{(l)}$ denotes the product of the $k-1$ partition functions associated with the rooted trees with cavity fields $g_{l,j}$ ($1\leq j\leq k-1$) and $\Xi^{(l)}$ ($1\leq l\leq k-1$) the partition function of the rooted tree obtained by connecting a new site $l$ to the $k$ rooted trees corresponding to $\Upsilon^{(l)}$ and $\Xi^{(l-1)}$.

We can write for one site addition
\begin{equation}
\frac{\Xi^{(l)}}{\Xi^{(l-1)}\Upsilon^{(l)}}=\exp [-m\mu\Delta\hat\Omega_{\rm iter}(h_{l-1};g_{l,j})],
\end{equation}
and for the last edge addition
\begin{equation}
\frac{\Xi}{\Xi^{(r-1)}\Xi^{(r)}}=\exp [-m\mu\Delta\hat\Omega_{\rm edge}(h_{r-1},g_r)].
\end{equation}
The 1-rsb formula for $\overline{\langle n_0n_r\rangle_c^p}$ is therefore
\begin{equation}
\label{eq:n0nr}
\begin{split}
 &\mathcal{Z}^{-1}\int dh_0P(h_0)\left[\prod_{l=1}^{r-1}\left(\prod_{j_l=1}^{k-1}dg_{l,j_l}P(g_{l,j_l})\right)\right]dg_rP(g_r)\\
&\left[\left(\frac{\partial\langle n_r\rangle}{\partial h_{r-1}}\right)^t\left(\prod_{l=1}^{r-1}\partial_1 \hat h(h_{l-1};g_{l,j})\right)\frac{\partial h_0}{\partial h_0^{(c)}} \right]^p\\
&\left(\prod_{l=1}^{r-1}e^{-m\mu\Delta\hat\Omega_{\rm iter}(h_{l-1};g_{l,j})}\right)e^{-m\mu\Delta\hat\Omega_{\rm edge}(h_{r-1},g_r)}
\end{split}
\end{equation}
where the normalization $\mathcal{Z}$ is given by
\begin{equation}
\begin{split}
\mathcal{Z}=&\int dh_0P(h_0)\left[\prod_{l=1}^{r-1}\left(\prod_{j_l=1}^{k-1}dg_{l,j_l}P(g_{l,j_l})\right)\right]dg_rP(g_r)\\
&\left(\prod_{l=1}^{r-1}e^{-m\mu\Delta\hat\Omega_{\rm iter}(h_{l-1};g_{l,j})}\right)e^{-m\mu\Delta\hat\Omega_{\rm edge}(h_{r-1},g_r)}.
\end{split}
\end{equation}
To be complete, we also need to insert in the previous formulae the following identity defining the intermediate fields $h_l$,
\begin{equation}
1=\int\prod_{l=1}^{r-1}dh_l\delta\left(h_l-\hat h(h_{l-1};g_{l,j})\right).
\end{equation}

To determine the behavior of the correlation functions between sites
at distance $r$, we introduce two transfer matrices, corresponding respectively to the numerator and denominator of Eq.~(\ref{eq:n0nr}),
\begin{equation}
\begin{split}
T_n(h_{l-1},h_l)&=\int \prod_{j_l=1}^{k-1}dg_{l,j_l}P(g_{l,j_l})\partial_1\hat h(h_{l-1},g_{l,j_l})\\
&\delta\left(h_l-\hat h(h_{l-1};g_{l,j_l})\right)e^{-m\mu\Delta\hat\Omega_{\rm
iter}(h_{l-1};g_{l,j_l})} ,\\
T_d(h_{l-1},h_l)&=\int \prod_{j_l=1}^{k-1}dg_{l,j_l}P(g_{l,j_l})e^{-m\mu\Delta\hat\Omega_{\rm iter}(h_{l-1};g_{l,j_l})}\\
&\delta\left(h_l-\hat h(h_{l-1};g_{l,j_l})\right)
.\\
\end{split}
\end{equation}

Finally, calling respectively $\lambda_{n}^{(p)}, \lambda_{d}$ the largest
eigenvalues of the matrices $(T_n)^p$ and $T_d$ we obtain that for large $r$
\begin{equation}
 \overline{\langle n_0 n_r\rangle_{c}^p}\sim \exp (-r/\xi_{p})
\end{equation}

with
\begin{equation}
\xi_{p}=-\frac{1}{\ln (|\lambda_n^{(p) }/\lambda_d|)}.
\end{equation}

Notice that all this discussion of stability of the 1-rsb solution is not just
academic. Indeed simulations performed on the BM with the distribution $P$ of the fields $g_{l,j}$ generated by population dynamics show that, at a fixed chemical potential $\mu>\mu_d$, the correlation length $\xi_{2}(m)$ increases when $m$ is decreased from $m=1$ to 0. In addition, the critical length $1/\ln 2$ is reached at some finite value of $m$, $m_c<m_s$, indicating that the description of metastable states corresponding to $m<m_c$ requires to break the replica symmetry beyond one step.

The limits of the 1-rsb approach will be discussed in much more details in the following section devoted to the $\mu=\infty$ limit where it is shown that the $[((d-1)\infty)\times((d-1)\infty)]$ transfer ``matrices'' (the $\infty$ stands for the continuum range of the fields $h_l$ so the matrices are actually operators) reduce to finite $[(d-1)d\times (d-1)d]$ matrices whose eigenvalues can be computed without resorting to the population dynamics.

\section{Close-packing limit}
\label{sec:5}

The zero temperature limit of the cavity method, which corresponds in lattice glasses to the $\mu=\infty$ limit, has received particular attention \cite{MezardParisi03}, both because of the simplifications it allows and because of its applications to optimization problems. For lattice glasses, the corresponding optimization problem, called the {\it close-packing problem}, consists in finding, for given lattice and packing constraint, the largest achievable particle density. We will obtain the solution of this problem as a result of our study.

\subsection{One-step rsb Ansatz}
\label{subsec:5:1}

We first consider the close-packing problem as a limiting case of the previous considerations; thus for $\mu\to\infty$, the single rs equations (\ref{eq:recursion1}) and (\ref{eq:recursion2}) simplify to
\begin{align}
\label{eq:reccp1}
a_0&=\max \left(0,1-\sum_{j=1}^ka_j+\max_{1\leq j\leq k}b_j\right),\\
\label{eq:reccp2}
b_0&=\max \left(0,1-\sum_{j=1}^ka_j\right).
\end{align}
The advantage of this limit is that we can resort to the exact Ansatz
\begin{equation}
\begin{split}
P(a,b)=&p_e\delta (a)\delta (b)+p_u\delta (a-1) \delta
(b-1)\\
&+(1-p_e-p_u) \delta (a-1) \delta (b).
\end{split}
\end{equation}

Note that the simple form obtained here results from our appropriate
choice of the local fields. Other choices may lead to a similar Ansatz
but with more than three spikes. This minimal number of three is related
to the three ``degrees of freedom'' of our model, as appeared
clearly when we needed three conditional partition functions. (In
general the number of spikes will be the minimum necessary number of local fields plus one).

This Ansatz is certainly the only one with integer fields, and at this
stage it is not obvious why we should not consider other solutions of Eq.~(\ref{eq:reccp1}-\ref{eq:reccp2}) with non integer fields. However the reason to take integer fields
appears clearly when working directly at $\mu=\infty$. Indeed, the
local fields have then a simple interpretation in term of the number
of particles and must therefore be integers. To see why, go back to the recursion on rooted trees and note $N_i^{(e)}$, $N_i^{(u)}$ and 
$N_i^{(s)}$ the numbers of particles of a (finite) rooted tree when its root node $i$ is empty ($e$), occupied but the constraint unsaturated ($u$)
and finally occupied and the constraint saturated ($s$),
i.e., the root site has $\ell$ neighboring particles. Considering as
before the $\ell =1$ case, we have
\begin{equation}
\begin{split}
N^{(e)}_0&=\sum_{j=1}^k\left(N^{(e)}_j+N^{(u)}_j+N^{(s)}_j\right),\\
N^{(u)}_0&=1+\sum_{j=1}^kN^{(e)}_j,\\
N^{(s)}_0&=1+\max_{1\leq j\leq k}N^{(u)}_j\sum_{p\neq j_{\rm max}}N^{(e)}_p
\end{split}
\end{equation}
where $N^{(u)}_{j_{\rm max}}\equiv\max_{1\leq j\leq k}N_j^{(u)}$.
Obviously, this is nothing but the corresponding $\mu\to\infty$ limit of the Eq.~(\ref{eq:xie}-\ref{eq:xis}) with $\Xi^{(a)}\sim\exp\left(\mu N^{(a)}\right)$, $a=e,u,s$. The corresponding local fields are
\begin{equation}
\begin{split}
a_i&=\max \left(N^{(e)}_i,N^{(u)}_i,N^{(s)}_i\right)-N^{(e)}_i,\\
b_i&=\max \left(N^{(e)}_i,N^{(u)}_i\right)-N^{(e)}_i.
\end{split}
\end{equation}
It is now clear that we can only have $a_i,b_i\in\{0,1\}$ with in
addition $b_i\leq a_i$. Moreover, one has a simple interpretation of the three spikes.

Plugging the Ansatz in the general 1-rsb cavity equation Eq.~(\ref{eq:cavity1-rsb}) and taking $y\equiv \lim_{\mu\to\infty}\mu m$ as breaking parameter  we get
\begin{align}
\label{eq:1-rsbCP1}
p_e&=Z^{-1}\left(1-p_e^k-kp_e^{k-1}p_u\right),\\
\label{eq:1-rsbCP2}
p_u&=Z^{-1}p_e^ke^y,\\
Z&=1+(e^y-1)(p_e^k+kp_e^{k-1}p_u).
\end{align}
These equations are in fact very simple and can be found following
the principle that for $\mu=\infty$ a particle must be present whenever it is allowed. So in terms of the state of the root, the merging of rooted graphs gives:
\begin{equation}
\label{eq:compositionrules}
\begin{split}
&\underbrace{e+\dots+e}_k \to u\\
&u+\underbrace{e+\dots+e}_{k-1} \to s\\
& {\rm all\ other\ combinations} \to e
\end{split}
\end{equation}
which simply means that $k$ empty sites lead to an occupied unsatured site, $k-1$ empty sites with one occupied unsatured leads to an occupied saturated sites and all other cases yield an empty site. Now call $p_e$, $p_u$ and $p_s$ the probabilities to be respectively in states $e$, $u$ and $s$; the three rules translate into three self-consistent equations
\begin{equation}
\begin{split}
p_u&\propto p_e^ke^y\\
p_s&\propto kp_up_e^{k-1}e^y\\
p_e&\propto 1-p_e^k-kp_up_e^{k-1}
\end{split}
\end{equation}
which are exactly the 1-rsb equations (\ref{eq:1-rsbCP1}-\ref{eq:1-rsbCP2}) for $\mu=\infty$ when the normalization $p_e+p_u+p_s=1$ is taken into account. The reweighting factor $e^y$ is introduced each time the recursion adds a particle since in this case we have a ``density shift'' $\Delta N=1$.

As before, the equilibrium value of the grand potential is given by the maximum at $y=y_s$ of $\phi (y)$ with here
\begin{equation}\label{phiy}
\begin{split}
-y\phi (y)=&\ln \left[ 1+(e^y-1)(p_e^{k+1}+(k+1)p_e^kp_u) \right]\\
&-\frac{k+1}{2}\ln \left[ 1+(e^{-y}-1)((1-p_e)^2-p_u^2)\right].
\end{split}
\end{equation}

\begin{figure}
\resizebox{0.5\textwidth}{!}{
\centering
\rotatebox{-90}{
\includegraphics{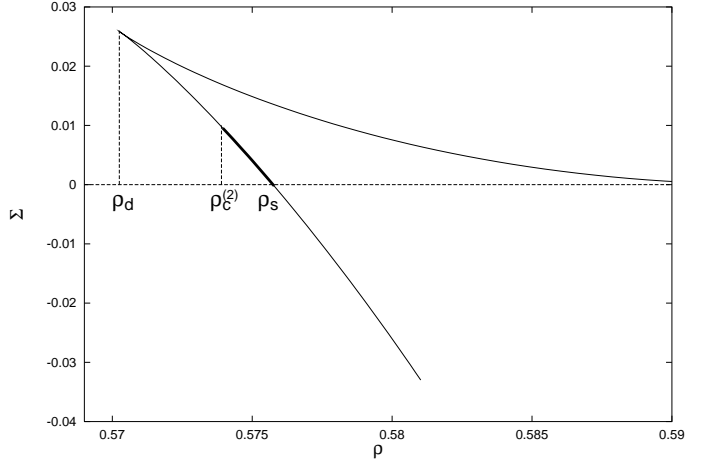}
}
}	
\caption{Complexity curve $\Sigma(\rho)$ obtained from the 1-rsb Ansatz for the BM in the close-packing limit $\mu=\infty$; its slope corresponds to $-y$. Four parts of the curve can be distinguished. First, a negative part ($y>y_s$) due to the contribution of untypical graphs with few frustrating loops. For $\rho_c^{(2)}<\rho<\rho_s$, i.e., $y_c^{(2)}<y<y_s$ (bold part), the 1-rsb Ansatz is stable and $\rho_s$ where $\Sigma=0$ gives the close-packing density. The complexity curve for $\rho<\rho_s$ corresponds to metastable states; it is no longer correctly described by the 1-rsb Ansatz for $\rho<\rho_c^{(2)}$ ($y<y_c^{(2)}$). Finally, we obtain for the fourth part an unphysical non concave branch.}
\label{fig:complexity}
\end{figure}

The configurational entropy $\Sigma(\rho)$ as a function of the density $\rho$ is defined by the parametrized curve
\begin{align}
\Sigma (y)&=y^2\partial_y \phi (y),\\
\rho (y)&=-\partial_y [y\phi (y)].
\end{align}
Note that since $\partial_y\phi(y_s)=0$, we have for the solution of the close-packing problem
\begin{equation}
\rho_\infty=\rho(y_s)=-\phi(y_s).
\end{equation}
The complexity $\Sigma$ has a maximum at $\rho_d=\rho(y_d)$ where $y_d$ is given by $\partial^2_y[y\phi(y)]=0$. The curve displays a non concave part for $y<y_d$ which may not have any physical interpretation; anyway, we will see that this unexpected part belongs to a region of the parameter $y$ where the results of a 1-rsb calculation are unreliable.
 In Fig.~\ref{fig:complexity}, we present the complexity curve for the BM. The close-packing densities of various models are presented in Fig.~\ref{fig:table}.

\begin{figure}
\centering
\begin{tabular}{c|c|c|c|c}
 $\ell$ & $k$ & $\rho_s$ & $\rho_d$ & $\rho_c^{(2)}$ \\
\hline \hline
1 & 2 & 0.575742 & 0.5703 & 0.5739 \\
\hline
1 & 3 & 0.517288 & 0.5097 & 0.5159 \\
\hline
1 & 4 & 0.473384 & 0.4646 & 0.4728 \\
\hline
1 & 5 & {\it 0.438382} & 0.4288 & frsb \\
\hline
2 & 2 & 0.735050 & 0.7302 & 0.7337 \\
\hline
2 & 3 & 0.636187 & 0.6256 & 0.6223 \\
\hline
2 & 4 & 0.573723 & 0.5606 & 0.5701 \\
\hline
2 & 5 & 0.527301 & 0.5129 & 0.5247 \\
\hline
3 & 3 & 0.776695 & 0.7748 & 0.7682 \\
\hline
3 & 4 & 0.680316 & 0.6660 & 0.6755 \\
\hline
3 & 5 & 0.617160 & 0.6001 & 0.6123 \\
\hline
4 & 4 & 0.805338 & 0.7945 & 0.8033 \\
\hline
4 & 5 & 0.713982 & 0.6972 & 0.7088 \\
\hline
5 & 5 & 0.826487 & 0.8140 & 0.8245 \\
\hline
\end{tabular}
\caption{Close-packing densities $\rho_\infty=\rho_s$ for random regular graphs of connectivity $k+1$ ($k=5$ approximates the three dimensional cubic space) where each particle can have no more than $\ell$ neighboring particles. We indicate the dynamical density $\rho_d$ where the complexity is maximum; any local algorithm trying to determinate $\rho_s$ will stay in the region where $\rho<\rho_d$. We emphasize however that this value is only a 1-rsb approximation (possibly an upper bound) which we have shown to be wrong due to the instability of second kind toward further rsb; $\rho_c^{(2)}$ gives the value of the density where this instability occurs and thus provides a lower bound for the correct $\rho_d$. Note that for $\ell=1$, $k=5$ even the equilibrium density $\rho_s$ is not correctly described by an 1-rsb Ansatz.}
\label{fig:table}
\end{figure}

\subsection{Stability of the 1-rsb Ansatz}
\label{subsec:5:2}

\begin{figure}
\resizebox{0.5\textwidth}{!}{
\rotatebox{-90}{
  \includegraphics{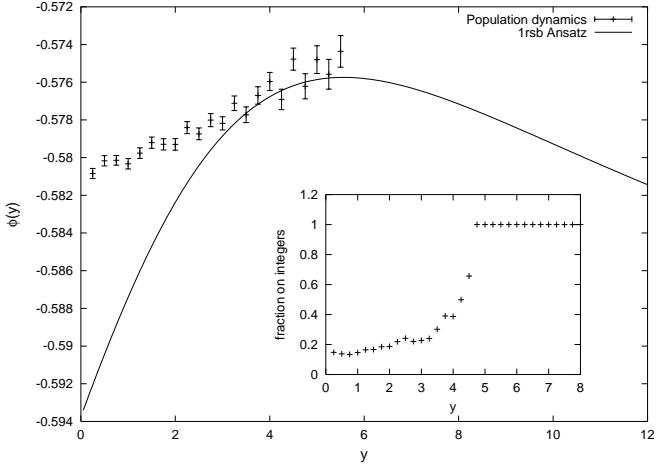}
}
}
\caption{$\phi (y)$ for the BM ($\ell=1$, $k=2$) in the close-packing limit $\mu=\infty$. The bold line is the result of the 1-rsb Ansatz. Its maximum at $y_s\simeq 5.56$ gives the close-packing density $\rho_s=-\phi(y_s)\simeq 0.5757$. The points with error bars were obtained with the population dynamics algorithm after $1000$ iterations of a population of $10000$ fields. We clearly obtain two different results for low $y$. Note that when $y$ increase, error bars increases, due to larger and larger reweighting factors, making $\phi(y)$ an average dominated by a few large terms only; this is why we do not display population dynamics results for large $y$ where anyway we know that the 1-rsb Ansatz should be recovered. A numerical check of this point is provided by studying directly the distribution of the fields. In the inset, we show how the fraction of the population within $\delta=0.01$ of one of the three spikes predicted by the 1-rsb Ansatz evolves with $y$. We thus verify that when $y>y_c^{(2)}\simeq 5.06$, the population is entirely on the peaks, even though it was started with an arbitrary distribution.}
\label{fig:phi(y)}
\end{figure}

Having derived the $\phi(y)$ function analytically, it is interesting
to compare it with the output of the population dynamics algorithm. As
a first consistency check, it is observed that when the population is
started on the integer spikes, both approaches lead exactly to the
same result. However, considering the stability of the integer Ansatz
under population dynamics provides additional features. For large
$y$, $y\geq y_c^{(2)}$, it is found that even when starting with arbitrary
fields, the dynamics converges to the expected distribution on the integers; however
for $y\leq y_c^{(2)}$, this Ansatz is found to be unstable and the
population dynamics converges to a new continuous distribution,
corresponding to a greater $\phi(y)$, as displayed in Fig.~\ref{fig:phi(y)}.
This behavior looks puzzling at first sight since on the one hand we
know that the fields must be integer, and on the other hand the cavity
method is known to lead to a lower bound of the grand potential \cite{FranzLeone03}, so given
two different solutions for $\phi(y)$, we must choose the larger. The
explanation for this contradiction must be that the approximation we
used, namely the 1-rsb formalism, is not valid. Therefore, we expect that
further replica symmetry breaking occurs for the metastable states with
$y\leq y_c^{(2)}$, a situation that has been argued to be a generic feature
of discontinuous spin glasses \cite{MontanariRicci03}.

We now present how the exact value $y_c^{(2)}$ can be computed. As when
dealing with the liquid instability, two approaches are possible. We
can either resort to the stability of the local fields distribution
under the cavity recursion, which requires to place oneself in a
two-step cavity formalism, or we can stay at the one-step level and
consider diverging response functions. 
The $\mu\to\infty$ limit of the formalism described 
in Sect.~\ref{subsec:4:10} is a bit tricky because correlations are 
trivial in the $\mu=\infty$ limit: $\overline{\langle n_0
n_r\rangle_c}=0$, due to a total freezing within each state. The 
stability analysis based on response functions is still possible, but
must not rely on susceptibilities; this procedure is presented in Appendix~B. Here, we will adopt an equivalent approach based on the 2rsb formalism, following \cite{MontanariRicci03}.

\subsubsection{Two-step replica symmetry breaking cavity method}

To emphasize the generality of the discussion, let us consider a generic
1-rsb solution at infinite $\mu$ given by a field distribution
\begin{equation}
\label{eq:1-rsbt=0}
P^*(h)=\sum_{a=1}^dp_a\delta(h-h_a)\equiv \sum_{a=1}^dp_a\delta_a(h)
\end{equation}
peaked on $d$ cavity fields, each having $(d-1)$ components that are integer valued [$d=3$ for all $(k,\ell)$ lattice glasses]. The $p_a$ satisfy a relation analogous to Eq.~(\ref{eq:1-rsbCP1}) and (\ref{eq:1-rsbCP2}),
\begin{equation}
	\label{eq:rec1-rsbt=0}
p_a=\frac{1}{Z}\sum_{(b_1,\dots,b_k)\to a}p_{b_1}\dots p_{b_k}\exp [-y \Delta E_{b_1,\dots,b_k}].
\end{equation}
Note that the fields are such that there is no degeneracy,
i.e., $k$ parents in configurations $(b_1,\dots,b_k)$ lead to a child
whose configuration can only be $a$. 

At the two-step level, not only the configurations are
grouped into different states, requiring to consider a distribution
$P(h)$ over these states, but the states are themselves organized into
larger clusters, that is groups of states sharing some common properties. We therefore need to consider a probability
distribution $Q[P]$ over the probability distributions $P(h)$. It has
the following meaning: on a given site, the distribution $P_c(h)$ of
the cavity field inside a cluster $c$ must be taken from the
distribution $Q[P]$.

The 2-rsb cavity equations are obtained by generalizing the 1-rsb distribution $\mathcal{N}_N(\omega)\sim e^{Ny(\omega-\omega_0)}$ of the number of states with fixed $\omega$ to
\begin{equation}
\label{eq:distr2-rsb}
\mathcal{N}_N(\omega)\sim \int d\omega_1 d\omega_2 \delta (\omega-\omega_1-\omega_2)e^{Ny_1(\omega_1-\omega_0)}e^{Ny_2(\omega_2-\omega_1)}.
\end{equation}
Here $\omega$ is decomposed into $\omega_1+\omega_2$ with $\omega_1$ the grand potential of a cluster with respect to a reference $\omega_0$, and $\omega_2$ the grand potential of a state inside the cluster with respect to that of the cluster $\omega_1$. The hierarchical rsb scheme is here reflected by the similarity between the distributions of $\omega_1$ and $\omega_2$, $\mathcal{N}^{(1)}_N(\omega_1|\omega_0)\sim \exp [Ny_1(\omega_1-\omega_0)]$ and $\mathcal{N}^{(2)}_N(\omega_2|\omega_1)\sim \exp [Ny_2(\omega_2-\omega_1)]$. Starting from the distribution given by Eq.~(\ref{eq:distr2-rsb}) and following the lines of the derivation of the 1-rsb cavity equation described in Sect.~\ref{subsec:4:4}, we obtain the 2-rsb cavity equation
\begin{equation}
Q[P]=\frac{1}{Z}\int\prod_{j=1}^k\mathcal{D}P_jQ[P_j]\hat z[\{P_j\}]^{y_1/y_2}\delta
[P-\hat P[\{P_j\}]]
\end{equation}
where
\begin{equation}
\begin{split}
\hat P[\{P_j\}]=\frac{1}{\hat z[\{P_j\}]}\int&\prod_{j=1}^kdh_jP_j(h_j)\delta
(.-\hat h(\{h_j\}))\\
&\exp (-y_2\Delta \hat E(\{h_j\}))
\end{split}
\end{equation}
and
\begin{equation}
\hat z[\{P_j\}]=\int\prod_{j=1}^kdh_jP_j(h_j)\exp (-y_2\Delta\hat E(\{h_j\})).
\end{equation}
The Parisi parameters $y_1$ and $y_2$ have to be taken such that $y_1\leq y_2$. Note that in particular, for $y_1=0$ and $y_2=y$ ($y$ denotes the 1-rsb parameter), the formalism describes a non-factorized 1-rsb solution.

It is essential to understand how the 1-rsb Ansatz must be written in
this 2-rsb formalism. Two scenarios are indeed possible. Either the 1-rsb
states coincide with the 2-rsb states and there is just one trivial
2-rsb cluster, or the 1-rsb states coincide with the 2-rsb clusters and
2-rsb states reduce to single configurations.
Within the first scenario, the one-step corresponds to $Q=\delta [P-P^*]$ where $P^*=\sum_ap_a\delta _a$ is the one-step probability distribution of Eq.~(\ref{eq:1-rsbt=0}) while within the second scenario, the one-step corresponds to
$Q[P]=\sum_ap_a\delta [P-\delta_a]$.

Depending on which case we consider, we can have two possible kinds of instabilities, as first noted by Montanari and Ricci-Tersenghi \cite{MontanariRicci03}.
As in Sect.~\ref{subsec:4:10}, in the first case, the states gather into different clusters, while in
the second case new states appear within the old states which therefore
become clusters. A pictorial view is given in Fig.~\ref{fig:instabilities}.

\subsubsection{Instability of the first kind: aggregation of states}
\label{subsubsec:5:2:2}

The instability of first kind can be studied by considering an Ansatz of the form 
\begin{equation}
Q[P]=f[P-P^*]
\end{equation}
where $f$ is a functional with support around the null function. 
The instability is given by the eigenvalue of largest modulus $\Lambda_1(y)$ of the Jacobian matrix associated with Eq.~(\ref{eq:rec1-rsbt=0}). Here again, as we deal with random graphs we ignore the modulation instability $k|\Lambda_1|>1$ and focus on the glass instability  $\sqrt{k}|\Lambda_1|>1$. Different cases are observed as we vary the parameters $\ell$ and $k$ in our lattice glass models. In some cases the instability is absent and appears only asymptotically, i.e., we have $\sqrt{k}|\Lambda_1(y)|<1$ for all $y$ but $\sqrt{k}|\Lambda_1(y)|\to 1$ as $y\to \infty$; this happens on low connectivity graphs, e.g. for $k=2,3$ when $\ell=1$. At higher connectivities, we can define a critical $y_c^{(1)}$ such that $\sqrt{k}|\Lambda_1(y)|>1$ for $y>y_c^{(1)}$. Then we have to determine the relative position of $y_c^{(1)}$ with respect to $y_s$ giving the maximum of $\phi(y)$. Fig.~\ref{fig:naturel1} shows how $y_s$ and $y_c^{(1)}$ evolve with the connectivity $k+1$ for the case $\ell=1$. When $y_s<y_c^{(1)}$, as it is found for $k=4,5$ for $\ell=1$, the positive part of the complexity curve is unaffected and we can rely on our 1-rsb description for typical graphs. However, if $y_s>y_c^{(1)}$, as we find when $6\leq k\leq 25$, the 1-rsb treatment is not stable, and one should develop a higher order rsb formalism.

\subsubsection{Instability of the second kind: fragmentation of states}
\label{subsec:5:2:3}

To study the instability of second kind, we consider an Ansatz of the form 
\begin{equation}
Q[P]=\sum_ap_af_a [P-\delta_a]
\end{equation}
 where the $f_a$ have support around the null function. Since $P$ is necessarily a combination of the $\delta_e$, the argument of $f_a$
can be written as $\delta P\equiv P-\delta_a=\sum_{e\neq a}\epsilon
_e(\delta_e-\delta_a)$ and we evaluate the widening of the $f_a$ by computing $\langle \delta P \rangle _a\equiv \int
\mathcal{D}\delta P f_a(\delta P)$ to obtain the following relations:
\begin{equation}
\begin{split}
\langle \epsilon _e \rangle _a=&\frac{k}{Zp_a}\sum_{(b_1\dots b_k)\to
  a}p_{b_1}\dots p_{b_k}e^{(y_2-y_1)\Delta E_{b_1,\dots, b_k}}\\
&\sum_{c\neq b_1,
  (c, b_2,\dots, b_k)\to e}e^{-y_2\Delta E_{c,b_2,\dots, b_k}}\langle
\epsilon_c\rangle_{b_1}.
\end{split}
\end{equation}
Since we consider a local instability of the 1-rsb solution we take $y_1=y_2=y$.
Noting $\lambda_{a\to e}\equiv \langle \epsilon_e\rangle_a$ it reads
\begin{equation}
\label{eq:apparentlycomplicated}
\begin{split}
p_a\lambda_{a\to e}=&\frac{1}{Z}\sum_{(b_1,\dots, b_k)\to a}p_{b_1}\dots p_{b_k}\\
&\sum_{j,c\neq b_j, (b_1,\dots,c,\dots, b_k)\to e}e^{-y\Delta E_{b_1,\dots,c,\dots,b_k}}\lambda_{b_j\to c}.
\end{split}
\end{equation}

This notation emphasizes the relation with the point of view based on response functions developed in Appendix B:
 $p(e|a)\equiv p_a\lambda_{a\to e}$ is the probability to replace $a$ with $e$ by changing only one parent. Indeed, for $\mu\to\infty$, adding a small perturbating field is equivalent to changing a configuration $a$. The equation tells us how such a change propagates from a site to its neighbor.

As expected from Sect.~\ref{subsec:4:10}, the instability is described
by a $[d(d-1)]\times[d(d-1)]$ transfer matrix $T_{ad,bc}$, $\langle
\epsilon _e \rangle _a=k\sum_{b\neq  c}T_{ae,bc}\langle
\epsilon_c\rangle_b$. If $\Lambda_2$ is the eigenvalue of $T$ of
largest modulus, the one-step solution is stable provided
$k\Lambda_2\leq 1$ ($\Lambda_2$ is positive, due to Perron-Frobenius
theorem); here we need not square the eigenvalues since the first
moments $\langle \epsilon _e \rangle _a$ does not vanish due to the
positivity of the $\lambda_{a\to e}$ (since $\Lambda_2$ is positive 
the instability detected by $k\Lambda_2 =1$ is compatible with the 
underlying random lattice).
Note that the same principles are straightforwardly extendable to cases with fluctuating connectivity and/or quenched disorder; we would have to average over the different possible transfer matrices.

We present in Appendix C a simple method to obtain the transfer matrix $T$ and focus here on the results. Generically, it is found that $k\Lambda_2<1$ for large enough $y$, $y>y^{(2)}_c$. For instance for the BM, we obtain $y^{(2)}_c\simeq 5.061$. This instability is precisely the one detected by the population dynamics algorithm. To underline this point, we determine the fraction of the population within $\delta=.01$ of one the $d=3$ expected delta peaks after 1000 iterations. At $y=5.06$, this fraction is found to represent 99.2 \% of the total population composed of 50000 individuals, while for $y=5.07$ (and all larger values) it is found to be 100 \% as shown in the inset of Fig.~\ref{fig:phi(y)}. We can thus conclude that the onset of non integer fields in the population dynamics is a clear sign that the replica symmetry must be broken beyond one step.

When comparing $y^{(2)}_c$ with $y_s$, the value of $y$ for which $\phi (y)$ taken with the 1-rsb Ansatz is maximum, we find either $y^{(2)}_c<y_s$ indicating that the close-packing is 1-rsb or $y^{(2)}_c>y_s$ suggesting that it is full rsb. For a given $\ell$, we obtain that $y^{(2)}_c<y_s$ at low connectivity ($k\leq k_1$, with $k_1=4,8,10$ for respectively $\ell=1,2,3$ ). But things are not exactly that simple since for instance when $\ell=1$ and $k\geq k_2=19$, $y^{(2)}_c<y_s$ again.

\subsection{Nature of the close-packing}
\label{subsec:5:4}

\begin{figure}
\resizebox{0.5\textwidth}{!}{
\rotatebox{-90}{
  \includegraphics{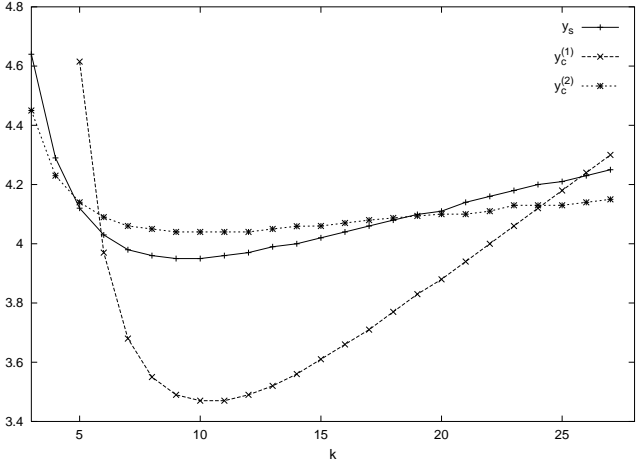}
}
}
\caption{For $\ell=1$ and various $k$, values of the parameters $y_s$, $y_c^{(1)}$ and $y_c^{(2)}$ giving respectively the equilibrium thermodynamics, the instability of the first kind (for $y> y_c^{(1)}$) and of the second kind  (for $y<y_c^{(2)}$). Only if $y_c^{(2)}<y_s<y_c^{(1)}$ is the close-packing limit correctly described by the 1-rsb Ansatz. That is the case for $k\leq 4$ and $k\geq 26$. The lines are only guides for the eyes (we omitted $y_c^{(1)}\simeq 7.0$ for $k=4$).}
\label{fig:naturel1}
\end{figure}

\begin{figure}
\resizebox{0.5\textwidth}{!}{
\includegraphics{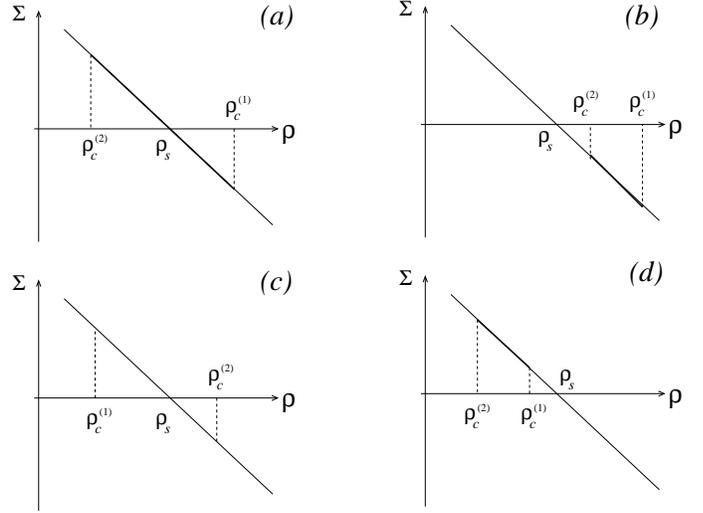}
}
\caption{Part of the complexity curve (bold line) correctly described by the 1-rsb Ansatz. For $\ell=1$, the case $(a)$ is realized for $k=2,3,4$; the models with $k=2$ (BM) and $k=3$ have in fact no instability of the first kind, i.e., the whole $\rho>\rho_c^{(2)}$ range is stable. Case $(b)$, realized for $k=5$, corresponds to a frsb close-packing, as well as case $(c)$ and $(d)$. In case $(c)$ no part of the 1-rsb complexity curve is correct; this happens for $6\leq k\leq 18$, the same conclusion holding for $19\leq k\leq 24$ where $\rho_c^{(1)}<\rho_s<\rho_c^{(2)}$ [case ($c'$), not represented here]. In case $(d)$, realized for $k=25$, some of the metastable states are indeed described by the 1-rsb Ansatz even if the close-packing is frsb.}
\label{fig:illustr}
\end{figure}

We now summarize the possible nature of the close-packing by taking into account both kinds of instabilities. Several situations can occur, corresponding to the different relative positions of $y_s$, the maximum of $\phi(y)$ giving the equilibrium properties with metastable states described by $y<y_s$, $y_c^{(1)}$ indicating an instability of the first kind for $y>y_c^{(1)}$, and $y_c^{(2)}$ indicating an instability of the second kind for $y<y_c^{(2)}$ (note the different directions). All possible combinations seem to be already contained in the $\ell=1$ model so we will detail it for connectivity ranging from $k=2$ to $k=26$; illustrations are provided by Fig.~\ref{fig:naturel1} and \ref{fig:illustr}.

Low connectivities $k=2,3,4$ make the best glass models since $y_c^{(2)}<y_s<y_c^{(1)}$, which means that the equilibrium state is 1-rsb and only less dense metastable states are frsb [case (a) of Fig.~\ref{fig:illustr}], a behavior similar to $p$-spin models \cite{MontanariRicci03}. When $k=5$, $y_s<y_c^{(2)}<y_c^{(1)}$ so the equilibrium state is frsb; however there exist some (but not all) untypical graphs with a 1-rsb phase [case (b) of Fig.~\ref{fig:illustr}]. When $6\leq k\leq 24$ the 1-rsb Ansatz is never stable since $y_c^{(1)}<y_c^{(2)}$; however the situation improves somewhat as $k$ increases, with $y_c^{(1)}<y_s<y_c^{(2)}$ for $6\leq k\leq 18$ (``doubly unstable'') and $y_c^{(1)}<y_c^{(2)}<y_s$ (``simply unstable'') for $19\leq k\leq 24$ [case (c) of Fig.~\ref{fig:illustr}]. The case $k=25$ is particularly interesting and justifies our study up to this large connectivity; indeed, we find $y_c^{(2)}<y_c^{(1)}<y_s$, i.e., the equilibrium state is frsb as well as the densest metastable states but there also exist metastable states with intermediate density which are well described by the 1-rsb approach [case (d) of Fig.~\ref{fig:illustr}]. Such a strange feature disappears for $k=26$ (and presumably for all higher $k$) where we recover the low connectivity situation with $y_c^{(2)}<y_s<y_c^{(1)}$.

\section{Dynamical and kinetic transitions}
\label{sec:6}

As explained in Sect.~\ref{subsec:4:9}, our lattice glass models undergo a {\it dynamical transition} at $\mu_d$ where the phase space breaks into exponentially many metastable states. Here ``dynamical'' means that the transition does not affect equilibrium properties, i.e., the grand potential is analytic at $\mu_d$ . However, despite its designation, the dynamical transition refers to a {\it static} property, in the sense that it describes a morphological change of the free energy landscape
and makes no reference to a particular kinetic rule. Kinetic transitions (also called ``dynamical arrests'' \cite{LawlorReagan02}) are associated with the freezing of some degrees of freedom of the system under specific local kinetic rules, and form the central concept of kinetically constrained models \cite{RitortSollich02}.

We now show that choosing the most natural local dynamical rules leads to several properties: (1) A kinetic transition at which a finite
fraction of particles is blocked takes place beyond a certain chemical
potential $\mu_k$. (2) The kinetic freezing at $\mu_k$ is a
distinct phenomenon from the ergodicity breaking occurring at $\mu_d$.
In particular it may happen that $\mu_k<\mu_d$ so that the dynamical
transition at $\mu_{d}$ is prevented by a kinetic arrest at $\mu_{k}$.

The easiest and more natural dynamical rule for our models (the one
that, by the way, one can use naturally to simulate it) is the standard
Monte Carlo, i.e., at each time step we pick at random a particle on a site $i$ and 
one of its neighboring site, $j$, and we move the particle onto $j$ provided that the geometrical constraint remains
satisfied after the move (site $j$ has no more than $\ell$ occupied neighboring site in addition to site $i$).
Furthermore, for simplicity, we focus on the simple $\ell=1$ model and instead of computing the exact kinetic transition $\mu_k$ as done in Ref.~\cite{ToninelliBiroli03} for the Kob-Andersen model on the Bethe lattice, we present a simpler calculation consisting in finding the onset of a particular blocked structure. This provides an upper bound $\mu_b>\mu_k$ for the kinetic transition, so that obtaining $\mu_d>\mu_b$ will be enough to prove that $\mu_d>\mu_k$ can occur.

\begin{figure}
\resizebox{0.5\textwidth}{!}{
\includegraphics{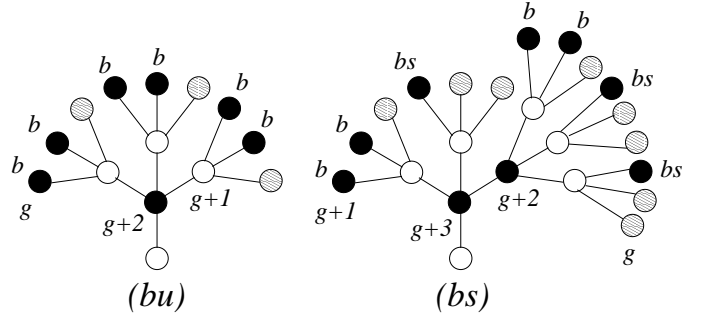}
}
\caption{Illustration of the blocked structure under consideration, with empty sites represented in white, sites occupied by a blocked particle in black and unspecified sites in grey. The structure is made of two kinds of blocked particles. The left figure shows how unsaturated blocked particles $(bu)$ at generation $g+2$ are prevented from moving to generation $g+1$ provided sites at generation $g+1$ have at least two blocked neighboring particles $b$ ($b$ meaning either $bu$ or $bs$). The right figure shows how saturated blocked particles $(bs)$ at generation $g+3$ are similarly blocked by blocked particles at generation $g+1$ and $g$. In this case a branch can be blocked either by a single $bs$ particle, or by two blocked particles.}
\label{fig:blocstruct}
\end{figure}

The blocked structure we consider is a percolating structure made of two kinds of blocked particles, either unsaturated or saturated (see Fig.~\ref{fig:blocstruct}). Rooted trees with an unsaturated and saturated blocked particle on the root are associated with the partition functions $\Xi^{(bu)}$ and $\Xi^{(bs)}$ respectively. The relation between the partition functions at different generations $g$ reads
\begin{equation}
\label{eq:blocked}
\begin{split}
\Xi^{(bu)}_{g+2}&=e^{\mu}\\
&\left[\left(\Xi_g\right)^k-\left( \Xi_g-\Xi^{(b)}_g\right)^k-k\Xi^{(b)}_g\left( \Xi_g-\Xi^{(b)}_g\right)^{k-1}\right]^k\\
\Xi^{(bs)}_{g+3}&=ke^{2\mu}\\
&\left[\left(\Xi_g\right)^k-\left( \Xi_g-\Xi^{(b)}_g\right)^k-k\Xi^{(bu)}_g\left(\Xi_g-\Xi^{(b)}_g\right)^{k-1}\right]^k\\
& \bigg [ (g \to g+1) \bigg ]^{k-1} 
\end{split}
\end{equation}
where $\Xi^{(b)}\equiv \Xi^{(bu)}+\Xi^{(bs)}$, and $\Xi\equiv \Xi^{(e)}+\Xi^{(u)}+\Xi^{(s)}$ is the total partition function (with notations of Sect.~\ref{subsec:3:1}).
As illustrated in Fig.~\ref{fig:blocstruct}, the equation for $\Xi^{(bu)}_{g+2}$ includes all configurations at generation $g$ [term $\left(\Xi_g\right)^k$] except those where no particle is blocked [$\left( \Xi_g-\Xi^{(b)}_g\right)^k$] and those where only one particle is blocked [$k\Xi^{(b)}_g\left(\Xi_g-\Xi^{(b)}_g\right)^{k-1}$]; in this case, the presence of {\it two} blocked particles on level $g$ is needed to exclude the possibility that a blocked particle on level $g$ could go at level $g+1$ and then to a different site at level $g$. The equation for $\Xi^{(bs)}_{g+3}$ has a similar interpretation.

In the liquid phase where the partition functions become independent of the generation $g$, we can consider the probability for a particle on the root to belong to the blocked structure,
\begin{equation}
p_b\equiv \frac{\Xi^{(b)}}{\Xi^{(u)}+\Xi^{(s)}}.
\end{equation}
From Eq.~(\ref{eq:blocked}) and (\ref{eq:xie}-\ref{eq:xis}), it is found to satisfy the self-consistent equation
\begin{eqnarray}
\label{eq:pb}
p_b=& \ \eta & \left[ 1-\left(1-\zeta p_b \right)^k-k\zeta p_b\left( 1-\zeta p_b\right)^{k-1}\right]^k \notag\\
&+ \theta & \bigg [ 1 - \left(1-\zeta p_b \right)^k-k\zeta \eta\left( 1 - \left(1-\zeta p_b \right)^k\right.\\
& & \left.-k\zeta p_b\left( 1-\zeta p_b \right)^{k-1}\right)^k\left( 1-\zeta p_b \right)^{k-1}\bigg ]^{2k-1} \notag
\end{eqnarray}
where $\zeta$, $\eta$ and $\theta$ depend on $\mu$ through the relations $
\zeta \equiv 1-e^{-\mu a_{\rm liq}}$, $\eta \equiv (e^{\mu b_{\rm
liq}}-1)/(e^{\mu a_{\rm liq}}-1)$ and $\theta \equiv (e^{\mu a_{\rm
liq}}-e^{\mu b_{\rm liq}})/(e^{\mu a_{\rm liq}}-1)$. At a given
connectivity $k$ we calculate $\mu_b$ as the lowest $\mu$ such that
Eq.~(\ref{eq:pb}) has a solution $p_b \in ]0,1]$. For $k\leq 5$ we
obtain $\mu_b>\mu_d$, i.e., the blocked structure considered does not
appear in the liquid phase. However, for $k=6$ we find $\mu_b\simeq
1.0$ which is lower than $\mu_d\simeq 2.7$; we can thus conclude for
sure that $\mu_k<\mu_d$, i.e., the kinetic freezing occurs while the
system is still in its liquid phase.

Let us conclude this section with some remarks on the difference
between the dynamical transition obtained from the cavity (or replica) computation
and the kinetic freezing that we studied in this section.
The kinetic freezing transition is clearly dependent on the dynamics
and is related to the fact that after a certain density there are no
more paths to go from one part of the configuration space to another. Of course, the allowed paths depend
on the local dynamics that has been chosen for the model. If one increases
the scale on which particles can move, for example
allowing particles jumps on next nearest neighbor or further, then the 
density $\rho_{k}$, at which the kinetic freezing takes place, is
expected to increase.
The thermodynamic limit, which does not play an important role for
kinetic freezing transitions, is instead very important for the dynamical transitions due to a
change of the morphology of the free energy landscape. In the latter case
when the number of sites becomes very large some bottlenecks in the
configuration space (that has to be used by any local dynamics)
shrinks. So, at the end, the configuration space breaks up in 
different ergodic components. 

\section{Discussion and conclusions}
\label{sec:7}
Some of the qualitative properties of our models 
such as the presence of a liquid phase at
low $\mu$ are easily understood.
Since our work concerns random graphs with no local disorder,
the equilibrium liquid phase is characterized by 
its homogeneity: the probability that a site is occupied 
is site-independent. Within the cavity framework, this is 
reflected by a unique value of the cavity fields
which describe in fact probabilities of particle occupation. 
Due to the tree-like structure of the lattices we consider, 
such probabilities can be computed recursively. However
such an approach is physically justified only if
the iterations converge to a unique fixed point regardless of the
initial conditions.

When $\mu$ is increased,
the liquid fixed point becomes unstable and
so some kind of equilibrium glassy phase must be present.
A clear evidence of the inadequacy of the liquid solution 
(i.e., of the Bethe-Peierls approximation) is the negative 
value of the liquid's entropy. In some cases, crystal phases 
can be constructed and we found these to be thermodynamically 
favored beyond a melting point $\mu_m$; however such ordered phases 
occur only on trees with carefully chosen boundary conditions. 
By focusing on \emph{random graphs}, we exclude this possibility, 
leaving as the only alternative the existence of an equilibrium glass phase.

The nature of the liquid to glass transition 
is discontinuous and of the one-step rsb type.
This means that the phase space first 
clusterizes into exponentially many non-ergodic components at 
some $\mu_d$, before a true equilibrium transition occurs 
at $\mu_s>\mu_d$. Both transitions are thermodynamical in 
nature but the transition at $\mu_d$ is called a ``dynamical transition'' 
to emphasize the fact that equilibrium properties are still given 
by the liquid solution when $\mu_d<\mu<\mu_s$, and 
only non-equilibrium properties are affected in this range.
More insights into these non-equilibrium effects where obtained 
by computing the configurational entropy $\Sigma(\omega)$ which 
gives the number $\mathcal{N}(\omega)$ of (metastable) clusters 
of configurations with a given grand potential $\omega$, through the 
relation $\mathcal{N}(\omega)=\exp[N\Sigma(\omega)]$ (with $N$ being
the number of sites of the graph).

Investigating the stability of the 1-rsb solution, we found that 
while the equilibrium configurations may be of 1-rsb type i.e., organized 
into distinct clusters, some metastable states can be associated with
full replica symmetry breaking (frsb), i.e., configurations organized into clusters 
themselves organized into smaller clusters, and so on. Our study 
of the $\mu=\infty$ case furthermore demonstrated that, depending 
on the model, the equilibrium state itself can exhibit frsb, 
and many different phase space structures are possible.

Finally, we showed that when using standard Monte Carlo dynamical
rule a kinetic freezing can occur at 
$\mu_k<\mu_d$, thus
preventing the dynamical transition which is due to a change of the free
energy landscape. In a general model, the freezing will therefore 
occur at $\mu_f=\min(\mu_k,\mu_d)$ and will be of a different 
nature, purely kinetic or truly thermodynamic, depending on 
which alternative is realized.
The overall picture for lattice glass models given
here is quite rich. A natural question is how
small loops modify (if at all) this picture; we expect
only small quantitative differences. 
A more fundamental and intriguing
issue is therefore what features survive when going
to finite dimensional lattices where the large scale
Euclidean structure could potentially change even
the qualitative aspects. 

The last point we would like to emphasize here is that even if 
our conclusions do not necessarily apply to real vitreous 
materials, our results should have an impact in optimization 
problems where random graphs arise naturally. Indeed, as a by-product 
of our study, we have determined the maximum densities and the
phase space structure of different close-packing problems. Our 
family of models provides a generalization to $\ell\geq 1$ of 
the hard-core model $\ell=0$ (called the {\em vertex cover} problem
in computer science and the  
{\em largest independent set} problem in the mathematics literature) 
with, in some cases, 1-rsb features much simpler than the frsb 
structure of the simplest ($\ell=0$) hard-core model. A full
understanding of these models could therefore constitute 
a first step toward a mathematical resolution of this fundamental 
but still unsolved problem \cite{AldousSteele03}.

\section*{Acknowledgements}
We thank Markus M{\"u}ller for helpful discussions, and in particular for explaining to us the method of Appendix C which he first developed in the context of lattice polymers \cite{MontanariMuller03}.

\section*{Appendix A}

In this appendix, we present the generalization of the formulae given for $\ell=1$ in the core of the paper to an arbitrary $\ell<k$, where $\ell$ is the maximum number of neighboring particles a given particle can have. It is straightforward to realize that here again, only three degrees of freedom are relevant ($d=3$) so that we also need only two-component local fields and the situation is very similar to $\ell=1$. The recursion relations are now
\begin{align}
\begin{split}
a_0=&\hat a(a_1,b_1,\dots,a_k,b_k)\\
=&\frac{1}{\mu}\ln \Big [1+e^{\mu (1-\sum_{j=1}^ka_j)}\\
&\left(1+\sum_{p=1}^l\sum_{1\leq i_1<\dots<i_p\leq k}\prod_{r=1}^p(e^{\mu b_{i_r}}-1)\right)\Big ],
\end{split}\\
\begin{split}
b_0=&\hat b(a_1,b_1,\dots,a_k,b_k)\\
=&\frac{1}{\mu}\ln \Big [1+e^{\mu (1-\sum_{j=1}^ka_j)}\\
&\left(1+\sum_{p=1}^{l-1}\sum_{1\leq i_1<\dots<i_p\leq k}\prod_{r=1}^p(e^{\mu b_{i_r}}-1)\right)\Big ].
\end{split}
\end{align}

In the $\mu\to\infty$ limit this becomes
\begin{equation}
\begin{split}
a_0&=\max \left(0,1-\sum_{j=1}^ka_j+\max_{1\leq i_1<\dots<i_l\leq k}\sum_{r=1}^lb_{i_r}\right),\\
b_0&=\max \left(0,1-\sum_{j=1}^ka_j+\max_{1\leq i_1<\dots<i_{l-1}\leq k}\sum_{r=1}^{l-1}b_{i_r}\right).
\end{split}
\end{equation} 
The Ansatz on integer peaks has exactly the same structure as for $\ell=1$,
\begin{equation}
\begin{split}
P(a,b)= &p_e\delta (a)\delta (b)+p_u\delta (a-1) \delta (b-1)\\
&+(1-p_e-p_u) \delta (a-1) \delta (b)
\end{split}
\end{equation}
but of course the relations between the $p_a$ need to be extended:
\begin{equation}
\begin{split}
p_e&=\frac{1}{Z}\left(1-\sum_{q=0}^l\binom{k}{q}p_u^qp_e^{k-q}\right),\\
p_u&=\frac{1}{Z}\sum_{q=0}^{l-1}\binom{k}{q}p_u^qp_e^{k-q}e^y,\\
Z&=1+(e^y-1)\sum_{q=0}^{l}\binom{k}{q}p_u^qp_e^{k-q}.
\end{split}
\end{equation}
Next we can write the corresponding $\phi(y)$ as
\begin{equation}
\begin{split}
-y\phi (y)=&\ln \left( 1+(e^y-1)\sum_{q=0}^l\binom{k+1}{q}p_1^qp_0^{k+1-q} \right)\\
&-\frac{k+1}{2}\ln \left( 1+(e^{-y}-1)[(1-p_0)^2-p_1^2]\right).
\end{split}
\end{equation}
The study of the stability follows the same principle as for $\ell=1$, the only difficulty being to correctly collect all the combinatorial factors. The matrix for the instability of the second kind is
\begin{equation}
T=
\begin{bmatrix}
0 & 0 & 0 & 0 & t_{01,20} & t_{01,21} \\
0 & 0 & t_{02,10} & 0 & t_{02,20} & t_{02,21} \\
0 & t_{10,02} & 0 & t_{10,12} & 0 & 0 \\
t_{12,01} & 0 & 0 & 0 & 0 & 0 \\
(k-l)e^{-y_2} & (k-l)e^{-y_2} & 0 & le^{-y_2} & 0 & 0\\
0 & 0 & l & 0 & 0 & 0
\end{bmatrix}
\end{equation}
with
\begin{equation}
\begin{split}
t_{01,20}&=\frac{1}{Z}\sum_{q=0}^{l-1}k\binom{k-1}{q}p_e^{k-q-2}p_u^qp_se^{y},\\
t_{01,21}&=\frac{1}{Z}\sum_{q=0}^{l-1}k\binom{k-1}{q-1}p_e^{k-q-1}p_u^{q-1}p_se^{y},\\
t_{02,10}&=\frac{1}{Z}(l+1)\binom{k}{l+1}p_e^{k-l-2}p_u^{l+1}e^{y},\\
t_{02,20}&=\frac{1}{Z}k\binom{k-1}{l}p_e^{k-l-2}p_u^lp_se^{y},\\
t_{02,21}&=\frac{1}{Z}k\binom{k-1}{l-1}p_e^{k-l-1}p_u^{l-1}p_se^{y},\\
t_{10,02}&=\frac{1}{Z}\sum_{q=0}^{l-1}(k-q)\binom{k}{q}p_e^{k-q}p_u^{q-1},\\
t_{10,12}&=\frac{1}{Z}\sum_{q=0}^{l-1}q\binom{k}{q}p_e^{k-q}p_u^{q-1},\\
t_{12,01}&=\frac{1}{Z}l\binom{k}{l}p_e^{k-l+1}p_u^{l-2}e^{y},\\
\end{split}
\end{equation}
where $p_s\equiv1-p_e-p_u$ and $Z=1+(e^{y_1}-1)\sum_{q=0}^{l}\binom{k}{q}p_u^qp_e^{k-q}$.

\section*{Appendix B}
In this appendix we shall show how the stability criterion at $\mu
=\infty $ derived in
the text can be obtained within a response function formalism. 

First let us note that as far as the fragmentation of the cluster is
concerned we can merely take the $\mu =\infty $ of the formalism 
introduced in Sect.~\ref{subsec:4:10}. Indeed in the close-packing
limit there is a total freezing within each state.
Hence, it is relatively easy to obtain the correlation length $\zeta_{2}$.
The computation of  $\overline{\langle n_in_j\rangle}$ can be done
using the transfer matrix:
\begin{equation}
T_{A,B}=\sum_{(A,a_{2}\dots,a_k)\to B}p_{a_{2}}\dots p_{a_{k}}\exp [-y \Delta E_{(A,a_{2},\dots,a_k)}],
\end{equation}
\begin{equation}
\overline{\langle n_0n_r\rangle}=\frac{P_0^tT^rP_0'}{P_*^tT^rP_*'}
\end{equation}
where $P_0^{t}$ and $P_0'$ are suitable vectors to impose that there
is a particle in $0$ and $r$.
The other vectors $P_*^t $ and  $P_*'$ read with the bracket notation:
\begin{equation}
\langle P_*|a\rangle=p_{a},\qquad \langle a|P_*'\rangle=\sum_{b{\rm\ compatible\ with\ }a} p_{b}.
\end{equation}

What is important to notice is that they are respectively the left and
right eigenvectors of $T$ with largest eigenvalues, $Z$, in module.
The properties of being eigenvectors can be checked but
it is fundamentally due to the fact that $p_{a}$
is the self-consistent solution of the iterative cavity equations which
can be written going from left to right or from right to left on the
1D chain. The fact that they correspond to the largest eigenvalue
in module is just due to the Perron-Frobenius theorem ($T$ is irreducible with
positive coefficient and the two eigenvectors have all the components
with the same sign).

Generically $T$ can be written in terms of its right and left eigenvectors:
\begin{equation}
T=\sum_{\alpha }\Lambda_{\alpha }|v_{\alpha }><w_{\alpha }|,\qquad
<w_{\beta }|v_{\alpha }>=\delta_{\alpha ,\beta },
\end{equation}
where the $\alpha =1$ corresponds to $P_*^t  $,  $P_*'$ (note
however that $P_*^t $ and  $P_*'$ are not normalized between
one another) and
$\Lambda_{1}=Z$.

Plugging this expression into the equations for the correlation function one
obtains generically that the correlation function reads:
\begin{equation}
\overline{\langle n_0n_r\rangle}=C+\sum_{i=2}^{d}D_{i}\left( \frac{\Lambda_{i}}{\Lambda_{1}}\right)^{r}
\end{equation}
where $C$ and $D_{i}$ are constants that can be expressed in terms of
the scalar product of the eigenvectors. In particular it can be shown
that $C=\rho^{2}$ as it should. Thus, the stability with respect to fragmentation of the cluster is
governed by the condition $k\left( \frac{|\Lambda_{2}|}{\Lambda_{1}}\right)^{2}<1$.

Finally, we want to show that this condition is the same as the one
given in Sect.~\ref{subsubsec:5:2:2}. 
Let us define the matrix $\tilde{T}$:
\begin{equation}
\tilde{T}_{A,B}=\frac{1}{Z} \left( T_{A,B}-p_{B}\sum_{C}T_{A,C}\right)
\end{equation}
This matrix has eigenvalues:$0,\frac{\Lambda_{2}}{\Lambda_{1}},\dots 
,\frac{\Lambda_{n}}{\Lambda_{1}}$. One can check this in different
ways. For example ${\rm Tr} \tilde{T}^{l}=\sum_{i=2}^{n}\left(
\frac{\Lambda_{i}}{\Lambda_{1}}\right)^{l}$ for any $l$, so the
eigenvalues are indeed the ones predicted. In terms of  $\tilde{T}$ the stability criterion is that 
$k$ times the square of the largest eigenvalue of $\tilde{T}$ has to
be less than one. This is exactly the criterion obtained in the text by studying the stability of the
iterative Eq.~(\ref{eq:1-rsbCP1}) for the 1-rsb glass (where the
 normalization $Z$ is considered as a function of the $p_a$). 

The second instability, the fragmentation of states, cannot be
obtained taking simply the $\mu\rightarrow \infty $ limit of the
formalism described in Sect.~\ref{subsec:4:10}. Because of the 
complete freezing within a state in the close packing limit 
the connected correlation functions are trivially zero. This, of
course, does not mean that there is no way to identify a correlation
length. Actually, one has just to resort to a definition of
correlation length in terms of response functions. 
Let us remark that the correlation function $\langle n_{0}
n_{r}\rangle_{c}$ can be written as:
\begin{equation}
p_{0}^{o}p_{0}^{e}\left(\langle n_{r}\rangle _{o}-\langle n_{r}\rangle _{e} \right)
\end{equation}
where $p_{0}^{o}$ and $p_{0}^{e}$ are respectively the probability
that the site $0$ is occupied or empty and the averages $\langle
\cdot \rangle _o$ and $\langle
\cdot \rangle _e$ are the averages conditioned on the events $n_{0}=1$
or $n_{0}=0$. At finite $\mu $ computing $\langle n_{0}
n_{r}\rangle_{c}$ or the response function $\left(\langle n_{r}\rangle
_{o}-\langle n_{r}\rangle _{e} \right)$ contains the same amount of
information as far as the correlation length is concerned. However at 
infinite $\mu $ there is total freezing within a state. As a
consequence the product $p_{0}^{o}p_{0}^{e}$ equals zero and one is
forced to extract the correlation length from the response functions.
Thus one has to study the change in probability for the occupation
variable at site $r$ when the value of occupation variable in $0$ is
changed. This gives rise naturally to the eigenvalue problem discussed
in Section \ref{subsec:5:2:3}. In that case the matrix entering in the 
eigenvalue problem tell us how the effect of replacing 
$a$ with $e$ by changing only one parent propagates from a site to its neighbor.
\section*{Appendix C}

We present here on the specific case of the lattice glass with $\ell=1$
how Eq.~(\ref{eq:apparentlycomplicated}) allows for a
straightforward study of the 1-rsb stability.
We start with the composition rules Eq.~(\ref{eq:compositionrules}). With the notations of this section, we have $d=3$ and the configurations are $a\in\{e,u,s\}$. Let us see for instance how the matrix elements corresponding to $\lambda_{s\to e}$ are computed. We have to ask how the configuration of one parent on the left hand-side of
\begin{equation}
\label{eq:initialreaction}
u+e+\dots+e\to s
\end{equation}
must be changed if we want to obtain a configuration $e$ instead of $s$. Begin with the parent $u$: according to the composition rules, if we replace it by $e$ it leads to a child $u$ so we do not retain this possibility; instead, if we replace it by $s$ we indeed obtain $e$. If we now keep $u$ but try to ``flip'' one of the $e$'s, we see that both replacements by $u$ or $s$ lead to a child $e$ as we wish. So we have a total of three cases to take into account, a situation that we summarize by writing the following ``reaction'' rules:
\begin{equation}
\label{eq:reactionequations}
\begin{split}
(e|s)&\leftarrow (s|u)+\underbrace{e+\dots+e}_{k-1}, \\
(e|s)&\leftarrow u+(u|e)+\underbrace{e+\dots+e}_{k-2}, \\
(e|s)&\leftarrow u+(s|e)+\underbrace{e+\dots+e}_{k-2}. \\
\end{split}
\end{equation}
Now recall that the initial ``reaction'' formula~(\ref{eq:initialreaction}) represents in fact the equation
\begin{equation}
p_s=Z^{-1}kp_up_e^{k-1}e^y
\end{equation}
 where $Z$ is a normalization constant, $kp_up_e^{k-1}$ is the probability to choose parents $(u,e,\dots,e)$ when each configuration $a$ is taken with its probability $p_a$, and the reweighting term $e^y$ is added each time the generated child is $u$ or $s$, i.e., corresponds to a new particle. Generalizing slightly the same principle, just by looking at the reaction equations (\ref{eq:reactionequations}), we write
\begin{equation}
\begin{split}
p(e|s)=&Z^{-1}kp(s|u)p_e^{k-1}+Z^{-1}k(k-1)p_up(u|e)p_e^{k-2}\\
&+Z^{-1}k(k-1)p_up(s|e)p_e^{k-2}.
\end{split}
\end{equation}
Here we put no reweighting since we generate an empty site $e$. Using the equations for the $p_a$, it can be simplified somewhat, and when expressed with the $\lambda_{a\to e}$ defined by $p(e|a)\equiv p_a\lambda_{a\to e}$, it becomes
\begin{equation}
\lambda_{s\to e}=[\lambda_{u\to s}+(k-1)\lambda_{e\to u}+(k-1)\lambda_{e\to s}]e^{-y}.
\end{equation}
The five other equations are obtained by following the same lines,
\begin{equation}
\begin{split}
\lambda_{u\to e}=&k\lambda_{e\to s}e^{-y},\\
\lambda_{u\to s}=&k\lambda_{e\to u},\\
\lambda_{s\to u}=&\lambda_{u\to e},\\
\lambda_{e\to u}=&Z^{-1}kp_sp_e^{k-2}\lambda_{s\to e}e^y,\\
\lambda_{e\to s}=&Z^{-1}kp_sp_e^{k-2}\lambda_{s\to u}e^y+Z^{-1}k(k-1)p_u^2p_e^{k-3}\lambda_{u\to e}e^y\\
&+Z^{-1}k(k-1)p_up_sp_e^{k-3}\lambda_{s\to e}e^y.\\
\end{split}
\end{equation}
These six linear relations define the matrix $T$; we then compute its eigenvalue $\Lambda_2$ as a function of $y$.  

\section*{Appendix D}
In this appendix we would like to show how the results obtained in the
text with the cavity method can be reproduced with replicas.
For an introduction to the replica method for random graphs with finite
connectivity see for example \cite{Monasson98}.

First, we introduce a function to encode in a compact way 
the conditional probability that the central site $i$ for a rooted sub-tree
is empty ({\it e}), occupied but the constraint unsaturated ({\it u})
and finally occupied and the constraint saturated ({\it s}),
i.e., the root site has $\ell$ neighboring particles.
We use the function $f (\tau )= C \exp (h\tau +g\tau^{2})$ with $\tau
=-1,0,1$ where $C,h,g$ are defined in such a way that
\begin{equation}
\begin{split}
f (0)&=\frac{\Xi_i^{(e)}}{\Xi_i^{(e)}+\Xi_i^{(u)}+\Xi_i^{(s)}},\\
f (1)&=\frac{\Xi_i^{(s)}}{\Xi_i^{(e)}+\Xi_i^{(u)}+\Xi_i^{(s)}},\\
f (-1)&=1-f (0)-f (1)=\frac{\Xi_i^{(u)}}{\Xi_i^{(e)}+\Xi_i^{(u)}+\Xi_i^{(s)}}.
\end{split}
\end{equation}
The relation with the fields $a,b$ can be easily established:
\begin{eqnarray}
\exp (-\mu a)&=&\frac{1}{1+\exp (g-h)+\exp (g+h)},\label{eq:agh}\\
\exp (-\mu b)&=&\frac{1}{1+\exp (g-h)}\label{eq:bgh}.
\end{eqnarray}

Now consider $n$ replicated systems for the same random graph and 
define the function $\psi (\vec{\tau })$:
\begin{equation}\label{def psi}
\psi (\vec{\tau })={\cal N}\overline{<\prod_{a=1}^{n}\exp (h_{i}^{(a)}\tau_{a} +g_{i}^{(a)}\tau_{a}^{2})>}
\end{equation}
where the overline means the average over the disorder 
(or, analogously, one can average over sites for a fixed random
graph), ${\cal N}$ is a normalization constant such that $\sum_{\vec{\tau }}\psi
(\vec{\tau })=1$ and $\vec{\tau }\equiv (\tau_1,\dots,\tau_n)$. 

For the same reasons as discussed in the text in the case of the cavity
method, the grand potential per site is given by:
\begin{align}
2\omega &=\lim_{n\rightarrow 0}\frac{1}{n}\left(-(k+1)G^{(n)}_{\rm edge}+2G^{(n)}_{\rm site}\right),\\
G^{(n)}_{\rm edge}&=\ln \sum_{\vec{\tau},\vec{\tau '}}\psi
(\vec{\tau })\psi (\vec{\tau '})C (\vec{\tau },\vec{\tau '}),\\
G^{(n)}_{\rm site}&=\ln \sum_{\vec{\tau_{i}},\vec{n_{o}}}\prod_{i=1}^{k+1}\psi
(\vec{\tau_{i}})C (\{ \vec{\tau_{i}}\},\vec{n_{0}})
\end{align}
where $C (\vec{\tau },\vec{\tau '})$ and $C (\{
\vec{\tau_{i}}\},\vec{n_{0}})$ enforce the geometrical constraint.
 
Differentiating $\omega $ with respect to $\psi
(\vec{\tau})$ we get the self-consistent equation
\begin{equation}
\psi
(\vec{\tau})={\cal N}\sum_{\vec{n_{0}},\vec{\vec{\tau }_{i}}}'\prod_{i=1}^{k}\psi
(\vec{\tau_{i}})C (\{ \vec{\tau_{i}}\},\vec{n_{0}})
\end{equation}
where the prime over $\Sigma$ means
that, for each value of $\vec{\tau }$, we sum only over those
$\vec{\tau_{i}},\vec{n_{0}}$ that give rise to $\vec{\tau }$.

Once the distribution $\psi
(\vec{\tau})$ is known all the observables can be computed. For
instance the density equals:
\begin{equation}\label{densityreplica}
\rho =\lim_{n\rightarrow 0}\frac{1}{n}\sum_{a,\vec{\tau_{i}},\vec{n_{o}}}n_{o}^{a}\prod_{i=1}^{k+1}\psi
(\vec{\tau_{i}})C (\{ \vec{\tau_{i}}\},\vec{n_{0}})
\end{equation}

The most general replica symmetric solution can be written:
\begin{equation}
\psi(\vec{\tau})=\int dhdg P (h,g) \prod_{a=1}^{n}\frac{\exp (h\tau_{a}
+g\tau^{2}_{a})}{1+\exp (g+h)+\exp (g-h)}
\end{equation}
where $P$ is a normalized probability distribution.
Because of the homogeneity of the random graph with fixed connectivity, 
the rs solution is particularly simple,
\begin{equation}
P (h,g)=\delta (h-h_{\rm liq})\delta (g-g_{\rm liq}).
\end{equation}

Using the relations (\ref{eq:agh}-\ref{eq:bgh}) between $h,g$ and $a,b$ one can rederive all the
liquid properties (density, free energy, entropy, \dots ) obtained in the main part of the paper with the cavity method.

Because of the homogeneity of the random graph, the 1-rsb solution, in which replicas are divided in $n/m$
groups (noted $c$) of $m$ replicas each, also becomes simple and reads
\begin{equation}
\psi(\vec{\tau})=\prod_{c=1}^{n/m}\int dh dg P (h,g)
\frac{\exp [\sum_{a\in c} (h\tau_{a}
+g\tau_{a}^{2})]}{[1+\exp (g+h)+\exp (g-h)]^{m}}.
\end{equation}
Plugging this expression into the general equation for
$\psi(\vec{\tau})$ and changing variables from $h,g$ to $a,b$ we get
back to the equation on $P (a,b)$, Eq.~(\ref{eq:cavity1-rsb}). Moreover using this 1-rsb expression of $\psi(\vec{\tau})$ 
one can easily obtain all the observables like the density, the free
energy and the complexity.

\bibliographystyle{prsty}

\bibliography{glasses,references}

\end{document}